\documentclass{JHEP3}%
\usepackage{amsfonts}
\usepackage{amsmath}
\usepackage{amssymb}
\usepackage{graphicx}
\usepackage{epsfig}

\title{An approach for the calculation of one-loop effective actions, vacuum
energies, and spectral counting functions}
\author{Wu-Sheng Dai and Mi Xie\\ Department of Physics, Tianjin University, Tianjin 300072, P.
R. China\\
LiuHui Center for Applied Mathematics, Nankai University \&
Tianjin University, Tianjin 300072, P. R. China \\
E-mail: \email{daiwusheng@tju.edu.cn}\\
E-mail: \email{xiemi@tju.edu.cn}}

\abstract{In this paper, we provide an approach for the
calculation of one-loop effective actions, vacuum energies, and
spectral counting functions and discuss the application of this
approach in some physical problems. Concretely, we construct the
equations for these three quantities; this allows us to achieve
them by directly solving equations. In order to construct the
equations, we introduce shifted local one-loop effective actions,
shifted local vacuum energies, and local spectral counting
functions. We solve the equations of one-loop effective actions,
vacuum energies, and spectral counting functions for free massive
scalar fields in $\mathbb{R}^{n}$, scalar fields in
three-dimensional hyperbolic space $H_{3}$ (the Euclidean Anti-de
Sitter space $AdS_{3}$), in $H_{3}/Z$ (the geometry of the
Euclidean BTZ black hole), and in $S^{1}$, and the Higgs model in
a $(1+1)$-dimensional finite interval. Moreover, in the above
cases, we also calculate the spectra from the counting functions.
Besides exact solutions, we give a general discussion on
approximate solutions and construct the general series expansion
for one-loop effective actions, vacuum energies, and spectral
counting functions. In doing this, we encounter divergences. In
order to remove the divergences, renormalization procedures are
used. In this approach, these three physical quantities are
regarded as spectral functions in the spectral problem.}
 \preprint{}
\begin{document}

\section{Introduction}

The main aim of this paper is to provide an approach for calculating one-loop
effective actions, vacuum energies, and spectral counting functions by
constructing their equations. The effective action plays an important role in
quantum field theory \cite{Avramidi}, which contains all the information of
quantized fields. The vacuum energy comes from the quantum fluctuation arising
from the uncertainty principle, which can be observed in, e.g., the Casimir
effect, and has consequences for the behavior of the universe on cosmological
scales \cite{CC,CC2,CC3}. The spectral counting function describes the number
of the eigenstates whose eigenvalues are smaller than a given number, which is
the core issue in the problem formulated by Kac as "Can one hear the shape of
a drum?" \cite{Kac}.

The regularized one-loop effective action $W_{s}$, the regularized vacuum
energy $E_{0}\left(  \epsilon\right)  $, and the spectral counting function
$N\left(  \lambda\right)  $ are global functions, i.e., they are not functions
of space coordinates. In practice, the global functions are very difficult to
calculate. An effective method for calculating the global functions is to
first calculate the corresponding local functions and then to achieve the
global ones from the local functions. The reason why it is relatively easy to
obtain the local function is that the local function has its own equation and
can be obtained by solving the equation. A typical example is the heat kernel:
the local heat kernel $K\left(  t;x,y\right)  $ can be obtained by solving the
heat equation, and the global heat kernel $K\left(  t\right)  $ can be
obtained by taking trace of $K\left(  t;x,y\right)  $.

Concretely, we first introduce the \textit{shifted} local versions of these
three global quantities: corresponding to $W_{s}$, $E_{0}\left(
\epsilon\right)  $, and $N\left(  \lambda\right)  $, we introduce the shifted
local one-loop effective action $W\left(  s;q;x,y\right)  $, the shifted local
vacuum energy $E_{0}\left(  \epsilon;q;x,y\right)  $, and the shifted local
spectral counting function $N\left(  \lambda;q;x,y\right)  $, respectively.
Then we construct the equations for these local ones; the global ones, $W_{s}
$, $E_{0}\left(  \epsilon\right)  $, and $N\left(  \lambda\right)  $, can be
obtained by taking trace of the corresponding shifted local ones with $q=0 $.
As a bridge, we construct the equation for the local Hurwitz zeta function
$\zeta\left(  s;q;x,y\right)  $ at first; $\zeta\left(  s;x,y\right)
=\zeta\left(  s;0;x,y\right)  $ is the known local zeta function
\cite{IM,M,Fulling}.

Some exact solutions of the shifted local one-loop effective action $W\left(
s;q;x,y\right)  $, the shifted local vacuum energy $E_{0}\left(
\epsilon;q;x,y\right)  $, and the local spectral counting function $N\left(
\lambda;x,y\right)  $ are solved from their equations in this paper. Such as a
free massive scalar field in $\mathbb{R}^{n}$, scalar fields in
three-dimensional hyperbolic space $H_{3}$ (the Euclidean Anti-de Sitter space
$AdS_{3}$) and in $H_{3}/Z$ (the geometry of the Euclidean BTZ black hole), a
scalar field in $S^{1}$, and the Higgs model in a $(1+1)$-dimensional finite
interval with the Dirichlet boundary condition. Based on the solved local
results, we, then, by taking trace, achieve the global ones, $W$, $E_{0}$, and
$N\left(  \lambda\right)  $. In order to obtain finite results,
renormalization procedures are used for removing the divergences. Moreover,
starting from a counting function, we calculate the eigenvalue spectrum of the
operator $D$ using the approach given in Ref. \cite{Ours}.

Besides exact solutions, we give a general discussion on series solutions,
which is a starting point for seeking approximate solutions. In order to
achieve an approximate solution, the first thing is to construct a proper
series expansion. In this paper, we construct the series expansions for
shifted local one-loop effective action $W\left(  s;q;x,y\right)  $, shifted
local vacuum energy $E_{0}\left(  \epsilon;q;x,y\right)  $, and local spectral
counting function $N\left(  \lambda;x,y\right)  $. Concretely, we first
construct the series expansion for the solution of general second-order
differential operators of Laplace type with local boundary conditions; such a
case is often related to the interaction case, such as gauge interactions, in
physical problems. Then, we construct the general form of the series
expansions for $W\left(  s;q;x,y\right)  $, $E_{0}\left(  \epsilon
;q;x,y\right)  $, and $N\left(  \lambda;x,y\right)  $. In finding the series
solutions, divergences are encountered and removed by renormalization procedures.

From a mathematical viewpoint, for an operator $D$ on a manifold $M$, the
character of $D$ and the geometry of $M$ are embodied in the spectrum
$\left\{  \lambda_{n}\right\}  $ determined by the eigenequation
\begin{equation}
D\phi_{n}=\lambda_{n}\phi_{n}.
\end{equation}
In principle, one can extract the information of $D$ and $M$ from the spectrum
$\left\{  \lambda_{n}\right\}  $. In modern researches, the study of spectrum
is often not through studying the eigenequation, but turns to the study of the
corresponding heat-type equation,
\begin{equation}
\partial_{t}\phi+D\phi=0,
\end{equation}
the wave-type equation,
\begin{equation}
\partial_{t}^{2}\phi+D\phi=0,
\end{equation}
the Schr\"{o}dinger-type equation,
\begin{equation}
i\partial_{t}\phi-D\phi=0,
\end{equation}
and, in principle, other equations in more general forms, by introducing
auxiliary variables (in the above mentioned three cases, the auxiliary
variable is the time $t$). For a given spectrum $\left\{  \lambda_{n}\right\}
$, different equations define different spectral functions, e.g., the spectral
function for heat-type equations is the fundamental solution
\begin{equation}
K\left(  t;x,y\right)  =\sum_{n}e^{-\lambda_{n}t}\phi_{n}\left(  x\right)
\phi_{n}^{\ast}\left(  y\right)
\end{equation}
(the heat kernel), for wave-type equations is the fundamental solution
\begin{equation}
\omega\left(  t;x,y\right)  =\sum_{n}e^{-i\sqrt{\lambda_{n}}t}\phi_{n}\left(
x\right)  \phi_{n}^{\ast}\left(  y\right)  ,
\end{equation}
and for Schr\"{o}dinger-type equations is the fundamental solution
\begin{equation}
h\left(  t;x,y\right)  =\sum_{n}e^{-i\lambda_{n}t}\phi_{n}\left(  x\right)
\phi_{n}^{\ast}\left(  y\right)  .
\end{equation}
For the spectrum $\left\{  \lambda_{n}\right\}  $, we can in principle
define\ other spectral functions which also embody the information of both $D$
and $M $. Effective actions, vacuum energies, and spectral counting functions
are all defined by the spectrum $\left\{  \lambda_{n}\right\}  $ and embody
the information of the operator $D$ and the manifold $M$, so they can serve as
spectral functions. That is to say, the local functions $N\left(
\lambda;q;x,y\right)  $, $W\left(  s;q;x,y\right)  $, $E_{0}\left(
\epsilon;q;x,y\right)  $, and $\zeta\left(  s;q;x,y\right)  $ are all spectral
functions for a spectral problem. These spectral functions are physical
meaningful and allow us to investigate the geometry of a manifold through
physical measures.

Many researches have been devoted to the study of the one-loop effective
action \cite{Vas2,Avramidi022,Avramidi023,Avramidi024,Avramidi025}. The zeta
function has many applications in spectrum problems \cite{Kbook}. Some
effective methods for calculating vacuum energies with the help of heat
kernels have been developed
\cite{EBK,EBK2,EBK3,EBK4,EBK5,Casimir,Casimir2,Casimir3,Casimir4}. There are
many studies on the heat kernel \cite{DK,DK2,DK3,DK4} and on its applications
\cite{Vic,Vic2,Vic3,Vic4}. For spectral counting functions, in mathematics,
the study sets off researches into spectral theory, with the idea of
recovering geometry of a manifold from the knowledge of the eigenvalues of a
differential operator \cite{Kac,Berger}. In physics, for example, one may seek
to reconstruct the shape of the universe from the eigenproblem \cite{ALSH}.
There are also experimental studies on spectral counting functions
\cite{Chaos}.

In section \ref{zcKdefrelations}, we construct the equations. In sections
\ref{one-loopaction}, \ref{vacuumenergy}, and \ref{spectralcountingfunction},
we first solve one-loop effective actions, vacuum energies, and spectral
counting functions for free massive scalar fields. Then we discuss the proper
series expansion for these three quantities. In sections \ref{H3}, \ref{S1},
and \ref{Higgs}, we solve the local and global one-loop effective actions,
vacuum energies, and counting functions for scalar fields in $H_{3}$,
$H_{3}/Z$, and $S^{1}$, and the Higgs model in a $(1+1)$-dimensional finite
interval, respectively. A discussion of spectra in such cases is also given in
these sections based on the result of the counting functions. The conclusions
are summarized in section \ref{conclusions}.

\section{Hurwitz zeta functions, one-loop effective actions, vacuum energies,
spectral counting functions, and heat kernels \label{zcKdefrelations}}

In this section, we construct the equations for one-loop effective actions,
vacuum energies, and spectral counting functions. For this purpose, we
introduce a shifted local regularized one-loop effective action $W\left(
s;q;x,y\right)  $, a shifted local regularized vacuum energy $E_{0}\left(
\epsilon;q;x,y\right)  $, and a shifted local spectral counting function
$N\left(  \lambda;q;x,y\right)  $, and generalize the local heat kernel
$K\left(  t;x,y\right)  $ to a \textit{shifted} local heat kernel $K\left(
t;q;x,y\right)  $. The corresponding unshifted global ones, $W_{s}$,
$E_{0}\left(  \epsilon\right)  $, $N\left(  \lambda\right)  $, and $K\left(
t\right)  $, can be obtained from these local ones by taking trace and setting
$q=0$.

To introduce these local functions, we start with the corresponding operators,
the zeta operator, the shifted counting operator, and the shifted heat kernel
operator; the shifted regularized one-loop effective action operator and the
shifted regularized vacuum energy operator can be directly achieved from the
zeta operator. The local functions are defined as the matrix elements of these
operators. The relations among these functions can be immediately obtained
from the definitions of the operators.

As a bridge, we first construct an equation for the local Hurwitz zeta
function, $\zeta\left(  s;q;x,y\right)  $ (The local zeta function also plays
an important role in many problems \cite{IM,M,Fulling}). Then, based on the
relations among $W\left(  s;q;x,y\right)  $, $E_{0}\left(  \epsilon
;q;x,y\right)  $, $N\left(  \lambda;q;x,y\right)  $, and $\zeta\left(
s;q;x,y\right)  $, we construct equations for the other three quantities.

Moreover, from the mathematical point of view, these local functions can be
regarded as spectral functions of a spectral problem.

\subsection{Definitions}

For an operator $D$ with spectrum $\left\{  \lambda_{n}\right\}  $, the global
heat kernel $K\left(  t\right)  =\sum_{n}e^{-\lambda_{n}t}$, the zeta function
$\zeta\left(  s\right)  =\sum_{n}\lambda_{n}^{-s}$, and the spectral counting
function $N\left(  \lambda\right)  =\sum_{n}\theta\left(  \lambda-\lambda
_{n}\right)  $, where $\theta\left(  x\right)  $ denotes the step function,
etc., can be viewed as various spectral functions of the spectral problem of
$D$. To construct the equations, we need the corresponding local functions. We
start with the corresponding operators. For an operator $D$, the zeta operator
is defined as $\mathbf{\zeta}=\left(  D+q\right)  ^{-s}$. Then the
global\ Hurwitz zeta function is $\zeta\left(  s;q\right)  =tr\mathbf{\zeta
}=\sum_{n}\left(  \lambda_{n}+q\right)  ^{-s}$. The one-loop effective action
operator reads $\mathbf{W}=\ln\sqrt{D}$, and the one-loop effective action is
$W=\operatorname*{tr}\mathbf{W}=\frac{1}{2}\ln\det D=%
%TCIMACRO{\dsum \nolimits_{n}}%
%BeginExpansion
{\displaystyle\sum\nolimits_{n}}
%EndExpansion
\ln\sqrt{\lambda_{n}}$. From the zeta operator, we can define a regularized
one-loop effective action operator: $\mathbf{W}_{s}=-\frac{1}{2}\tilde{\mu
}^{2s}\Gamma\left(  s\right)  \mathbf{\zeta}$, where $\tilde{\mu}$ is a
constant. The global regularized one-loop effective action is $W_{s}=tr\left.
\mathbf{W}_{s}\right\vert _{q=0}=-\frac{1}{2}\tilde{\mu}^{2s}\Gamma\left(
s\right)  \sum_{n}\lambda_{n}^{-s}$. From the zeta operator, we can also
define a shifted regularized vacuum energy operator: $\mathbf{E}_{0}=\frac
{1}{2}\tilde{\mu}^{2\epsilon}\left(  D+q\right)  ^{1/2-\epsilon}=\frac{1}%
{2}\left.  \mathbf{\zeta}\right\vert _{s=-1/2+\epsilon}$. Its trace with $q=0$
gives the regularized vacuum energy: $E_{0}\left(  \epsilon\right)  =tr\left.
\mathbf{E}_{0}\right\vert _{q=0}=\frac{1}{2}\tilde{\mu}^{2\epsilon}\left.
\zeta\left(  -1/2+\epsilon\right)  \right\vert _{q=0}$; a renormalized vacuum
energy $E_{0} $ can be directly obtained from $E_{0}\left(  \epsilon\right)  $.

The shifted counting operator is defined as $\mathbf{N}=\theta\left(
\lambda-\left(  D+q\right)  \right)  $. The spectral counting function reads
$N\left(  \lambda\right)  =tr\left.  \mathbf{N}\right\vert _{q=0}=\sum
_{n}\theta\left(  \lambda-\lambda_{n}\right)  $. We also introduce a shifted
heat kernel operator, $\mathbf{K}=e^{-\left(  D+q\right)  t}$. The trace of
$\mathbf{K}$ gives the shifted global heat kernel: $K\left(  t;q\right)
=tr\mathbf{K}=\sum_{n}e^{-\left(  \lambda_{n}+q\right)  t}$.

The local functions are defined as the matrix elements of the operators. The
local Hurwitz zeta function is the matrix element of the zeta operator,%
\begin{equation}
\zeta\left(  s;q;x,y\right)  =\left\langle x\left\vert \mathbf{\zeta
}\right\vert y\right\rangle =\sum_{n}\left(  \lambda_{n}+q\right)  ^{-s}%
\phi_{n}\left(  x\right)  \phi_{n}^{\ast}\left(  y\right)  .\label{zetadef}%
\end{equation}
$\zeta\left(  s;x,y\right)  =\zeta\left(  s;0;x,y\right)  $ is just the known
local zeta function \cite{IM,M,Fulling}. The shifted local regularized
one-loop effective action is the matrix element of the regularized one-loop
effective action operator,
\begin{equation}
W\left(  s;q;x,y\right)  =\left\langle x\left\vert \mathbf{W}_{s}\right\vert
y\right\rangle =-\frac{1}{2}\tilde{\mu}^{2s}\Gamma\left(  s\right)
\zeta\left(  s;q;x,y\right)  .\label{defW}%
\end{equation}
The shifted local vacuum energy is the matrix element of the shifted
regularized vacuum energy operator,%
\begin{equation}
E_{0}\left(  \epsilon;q;x,y\right)  =\frac{1}{2}\left\langle x\left\vert
\mathbf{E}_{0}\right\vert y\right\rangle =\frac{\tilde{\mu}^{2\epsilon}}%
{2}\zeta\left(  -\frac{1}{2}+\epsilon;q;x,y\right)  .\label{veo}%
\end{equation}
Note that $E_{0}\left(  x\right)  =E_{0}\left(  0;0;x,x\right)  $, the
unshifted diagonal case of $E_{0}\left(  \epsilon;q;x,y\right)  $, is just the
vacuum energy density \cite{Fulling}.

The matrix element of the shifted counting operator defines the shifted local
spectral counting function:%
\begin{equation}
N\left(  \lambda;q;x,y\right)  =\left\langle x\left\vert \mathbf{N}\right\vert
y\right\rangle =\sum_{n}\theta\left(  \lambda-\left(  \lambda_{n}+q\right)
\right)  \phi_{n}\left(  x\right)  \phi_{n}^{\ast}\left(  y\right)
;\label{Ndef}%
\end{equation}
the matrix element of the shifted heat kernel operator defines the shifted
local heat kernel:%
\begin{equation}
K\left(  t;q;x,y\right)  =\left\langle x\left\vert \mathbf{K}\right\vert
y\right\rangle =\sum_{n}e^{-\left(  \lambda_{n}+q\right)  t}\phi_{n}\left(
x\right)  \phi_{n}^{\ast}\left(  y\right)  .\label{Kdef}%
\end{equation}

For the case of $q=0$, $N\left(  \lambda;x,y\right)  =N\left(  \lambda
;0;x,y\right)  $ defines the local spectral counting function, and $K\left(
t;x,y\right)  =K\left(  t;0;x,y\right)  $ is the local heat kernel. We have
$K\left(  t;q;x,y\right)  =e^{-qt}K\left(  t;x,y\right)  $ and $N\left(
\lambda;q;x,y\right)  =N\left(  \lambda-q;x,y\right)  $. The trace of an
operator, taking the spectral counting function as an example, can be taken
as
\begin{equation}
N\left(  \lambda\right)  =tr\left.  \mathbf{N}\right\vert _{q=0}=\int
d^{n}x\sqrt{g}N\left(  \lambda;x,x\right)  =\sum_{n}\theta\left(
\lambda-\lambda_{n}\right)  =\sum_{\lambda_{n}<\lambda}1.
\end{equation}

\subsection{Relations}

Mathematically speaking, all the functions mentioned above are essentially
various spectral functions for a given operator $D$. The relations among them
can be formally deduced from their definitions. From eqs. (\ref{zetadef}),
(\ref{Ndef}), and (\ref{Kdef}), we can achieve the relations among
$\zeta\left(  s;q;x,y\right)  $, $N\left(  \lambda;q;x,y\right)  $, and
$K\left(  t;q;x,y\right)  $.

By the representation
\begin{equation}
D^{-s}=s\int_{0}^{\infty}d\lambda\frac{1}{\lambda^{s+1}}\theta\left(
\lambda-D\right)  ,
\end{equation}
we achieve%
\begin{equation}
\zeta\left(  s;q;x,y\right)  =s\int_{0}^{\infty}d\lambda\frac{1}{\lambda
^{s+1}}N\left(  \lambda;q;x,y\right)  ,\label{zetaN}%
\end{equation}
and by the representation
\begin{equation}
D^{-s}=\frac{1}{\Gamma\left(  s\right)  }\int_{0}^{\infty}dt\,t^{s-1}e^{-tD},
\end{equation}
we achieve%
\begin{equation}
\zeta\left(  s;q;x,y\right)  =\frac{1}{\Gamma\left(  s\right)  }\int%
_{0}^{\infty}dt\,t^{s-1}K\left(  t;q;x,y\right)  .\label{ZetaandK}%
\end{equation}
The corresponding inverse transformations can be obtained directly, e.g.,%
\begin{equation}
N\left(  \lambda;q;x,y\right)  =\frac{1}{2\pi i}\int_{-c-i\infty}^{-c+i\infty
}ds\frac{\lambda^{s}}{s}\zeta\left(  s;q;x,y\right)  .
\end{equation}
We then obtain the relation between $N\left(  \lambda;q;x,y\right)  $ and
$K\left(  t;q;x,y\right)  $:%
\begin{align}
K\left(  t;q;x,y\right)   &  =t\int_{0}^{\infty}d\lambda N\left(
\lambda;q;x,y\right)  e^{-\lambda t},\label{NKq}\\
N\left(  \lambda;q;x,y\right)   &  =\frac{1}{2\pi i}\int_{c-i\infty
}^{c+i\infty}dt\frac{e^{\lambda t}}{t}K\left(  t;q;x,y\right)  .\label{KNq}%
\end{align}

The relations among $N\left(  \lambda\right)  $, $K\left(  t\right)  $, and
$\zeta\left(  s\right)  $ can be immediately obtained from the\ corresponding
local relations:
\begin{equation}
\zeta\left(  s\right)  =s\int_{0}^{\infty}d\lambda\frac{N\left(
\lambda\right)  }{\lambda^{s+1}}=\frac{1}{\Gamma\left(  s\right)  }\int%
_{0}^{\infty}dt\,t^{s-1}K\left(  t\right)  .
\end{equation}
An in-depth discussion of the global heat kernel and the spectral counting
function has been provided in Refs. \cite{Kirsten,Kirsten2}.

Moreover, the relations among $\zeta\left(  s;q;x,y\right)  $, $W\left(
s;q;x,y\right)  $, $E_{0}\left(  \epsilon;q;x,y\right)  $, and $N\left(
\lambda;q;x,y\right)  $ can be obtained straightforwardly. For example, from
eqs. (\ref{defW}) and (\ref{zetaN}), we have
\begin{equation}
W\left(  s;q;x,y\right)  =-\frac{1}{2}\tilde{\mu}^{2s}\Gamma\left(
s+1\right)  \int_{0}^{\infty}d\lambda\frac{N\left(  \lambda;q;x,y\right)
}{\lambda^{s+1}}.
\end{equation}

\subsection{Equations}

\textit{The equation for Hurwitz zeta functions.} The equation for
$\zeta\left(  s;q;x,y\right)  $ can be constructed as
\begin{equation}
\left(  s-1\right)  \int^{q}dq\zeta\left(  s;q;x,y\right)  +\left(
D_{x}+q\right)  \zeta\left(  s;q;x,y\right)  =0,\label{deqofzeta}%
\end{equation}
with the condition $\lim_{t\rightarrow0}\left[  \int_{c-i\infty}^{c+i\infty
}ds\Gamma\left(  s\right)  t^{-s}\zeta\left(  s;0;x,y\right)  \right]
=i2\pi\delta\left(  x-y\right)  $, where the order of the integrate and the
limit cannot be exchanged. This is a partial integro-differential equation
\cite{AKZ}. Such an equation can be translated into a partial differential
equation by taking derivative with respect to $q$:%
\begin{equation}
D_{x}\frac{\partial}{\partial q}\zeta\left(  s;q;x,y\right)  +q\frac{\partial
}{\partial q}\zeta\left(  s;q;x,y\right)  +s\zeta\left(  s;q;x,y\right)  =0.
\end{equation}
The reason why we use the local Hurwitz zeta function $\zeta\left(
s;q;x,y\right)  $ which can be viewed as a shifted local zeta function rather
than the local zeta function $\zeta\left(  s;x,y\right)  $ is that though for
the aim of constructing an equation for the zeta function, only a local zeta
function $\zeta\left(  s;x,y\right)  $ is sufficient, the equation for
$\zeta\left(  s;x,y\right)  $ is a recurrence differential equation,
\begin{equation}
D_{x}\zeta\left(  s;x,y\right)  -\zeta\left(  s-1;x,y\right)  =0,
\end{equation}
which is difficult to deal with. If we adopt the local Hurwitz zeta function,
then by the relation $\frac{\partial}{\partial q}\zeta\left(  s;q;x,y\right)
=-s\zeta\left(  s+1;q;x,y\right)  $, we achieve eq. (\ref{deqofzeta}) instead
of the recurrence differential equation.

\textit{The equation for regularized one-loop effective actions.} The equation
for $W\left(  s;q;x,y\right)  $ can be obtained from eqs. (\ref{defW}) and
(\ref{deqofzeta}):%
\begin{equation}
\left(  s-1\right)  \int^{q}dqW\left(  s;q;x,y\right)  +\left(  D_{x}%
+q\right)  W\left(  s;q;x,y\right)  =0;\label{EqW}%
\end{equation}
with the condition $\lim_{t\rightarrow0}\left[  \int_{c-i\infty}^{c+i\infty
}dsW\left(  s;0;x,y\right)  /\left(  \tilde{\mu}^{2s}t^{s}\right)  \right]
=-i\pi\delta\left(  x-y\right)  $. Taking derivative with respect to $q$ gives
the corresponding partial differential equation:
\begin{equation}
D_{x}\frac{\partial}{\partial q}W\left(  s;q;x,y\right)  +q\frac{\partial
}{\partial q}W\left(  s;q;x,y\right)  +sW\left(  s;q;x,y\right)
=0.\label{EqWdff}%
\end{equation}
By the way, if we start with an unshifted local regularized one-loop effective
action $W\left(  s;x,y\right)  =W\left(  s;0;x,y\right)  $, we will obtain a
recurrence differential equation:
\begin{equation}
D_{x}W\left(  s;x,y\right)  -\left(  s-1\right)  \tilde{\mu}^{2}W\left(
s-1;x,y\right)  =0.
\end{equation}

\textit{The equation for regularized vacuum energies. }The equation for the
shifted local vacuum energy can be obtained directly from eqs. (\ref{veo}) and
(\ref{deqofzeta}):%
\begin{equation}
\frac{3}{2}\int^{q}dqE_{0}\left(  \epsilon;q;x,y\right)  -\epsilon\int%
^{q}dqE_{0}\left(  \epsilon;q;x,y\right)  -\left(  D_{x}+q\right)
E_{0}\left(  \epsilon;q;x,y\right)  =0,
\end{equation}
with the condition $\lim_{t\rightarrow0}\left[  \int_{c-i\infty}^{c+i\infty
}d\epsilon\Gamma\left(  -1/2+\epsilon\right)  t^{1/2-\epsilon}\tilde{\mu
}^{-2\epsilon}E_{0}\left(  \epsilon;0;x,y\right)  \right]  =i\pi\delta\left(
x-y\right)  $. Taking derivative with respect to $q$ gives the corresponding
partial differential equation:%
\begin{equation}
D_{x}\frac{\partial}{\partial q}E_{0}\left(  \epsilon;q;x,y\right)
+q\frac{\partial}{\partial q}E_{0}\left(  \epsilon;q;x,y\right)  -\frac{1}%
{2}E_{0}\left(  \epsilon;q;x,y\right)  +\epsilon E_{0}\left(  \epsilon
;q;x,y\right)  =0.\label{EqofveDeq}%
\end{equation}
The regularized vacuum energy can be achieved by taking trace with $q=0$.

\textit{The equation for spectral counting functions. }The equation for
$N\left(  \lambda;x,y\right)  $ can be obtained from the equation of
$\zeta\left(  s;q;x,y\right)  $ with $q=0$ by using the relation
(\ref{zetaN}):%
\begin{equation}
\int^{\lambda}d\lambda N\left(  \lambda;x,y\right)  +\left(  D_{x}%
-\lambda\right)  N\left(  \lambda;x,y\right)  =0\label{dieqofN}%
\end{equation}
with the condition $\lim_{t\rightarrow0}\left[  \int_{0}^{\infty}d\lambda
N\left(  \lambda;x,y\right)  te^{-\lambda t}\right]  =\delta\left(
x-y\right)  $. Eq. (\ref{dieqofN}) is a partial integro-differential equation.
Taking derivative with respect to $\lambda$ to both sides of eq.
(\ref{dieqofN}) will give the corresponding partial differential equation:
\begin{equation}
D_{x}\frac{\partial}{\partial\lambda}N\left(  \lambda;x,y\right)
-\lambda\frac{\partial}{\partial\lambda}N\left(  \lambda;x,y\right)
=0.\label{eqofNdiff}%
\end{equation}
Defining a local state density $\rho\left(  \lambda;x,y\right)  \equiv
\frac{\partial}{\partial\lambda}N\left(  \lambda;x,y\right)  $, we achieve
\begin{equation}
D_{x}\rho\left(  \lambda;x,y\right)  =\lambda\rho\left(  \lambda;x,y\right)  .
\end{equation}
The global state density $\rho\left(  \lambda\right)  $ can be obtained by
taking trace of $\rho\left(  \lambda;x,y\right)  $.

\section{Solutions of local and global one-loop effective actions:
Renormalization \label{one-loopaction}}

\subsection{The free-field solution}

In this section, we first solve the shifted local regularized one-loop
effective action for a free massive scalar field in $\mathbb{R}^{n}$ from eq.
(\ref{EqW}). In this case, $D_{0}=-\nabla^{2}+m^{2}$, where $m$ is the mass.
The solution of eq. (\ref{EqW}) reads%
\begin{equation}
W_{0}\left(  s;q;x,y\right)  =-\frac{\tilde{\mu}^{2s}}{\left(  4\pi\right)
^{n/2}}\left(  \frac{2\sqrt{m^{2}+q}}{\left\vert x-y\right\vert }\right)
^{n/2-s}K_{-n/2+s}\left(  \sqrt{m^{2}+q}\left\vert x-y\right\vert \right)
,\label{Wscalar}%
\end{equation}
where $K_{n}\left(  z\right)  $ is the modified Bessel function.

As a function of $x$ and $y$, $W_{0}\left(  s;q;x,y\right)  $ has a singular
point at $\left\vert x-y\right\vert =0$, or, $W_{0}\left(  s;q;x,y\right)  $
is analytic except for $\left\vert x-y\right\vert =0$. However, when seeking a
global one-loop effective action, what we concern is just the point
$\left\vert x-y\right\vert =0$ since the global one-loop effective action is
obtained through taking trace of the local one. This means that when achieving
a global one-loop effective action from the corresponding local one-loop
effective action, we need a renormalization procedure to remove the divergence.

In order to extract the divergence led by $\left\vert x-y\right\vert =0$, by
use of the expansion of $K_{\nu}\left(  z\right)  $ around $z=0$,
\begin{equation}
K_{\nu}\left(  z\right)  =\frac{1}{2}\Gamma\left(  \nu\right)  \left(
\frac{2}{z}\right)  ^{\nu}\sum_{p=0}^{\infty}\frac{\left(  z/2\right)  ^{2p}%
}{\left(  1-\nu\right)  _{p}p!}+\frac{1}{2}\Gamma\left(  -\nu\right)  \left(
\frac{z}{2}\right)  ^{\nu}\sum_{p=0}^{\infty}\frac{\left(  z/2\right)  ^{2p}%
}{\left(  \nu+1\right)  _{p}p!},\label{lim}%
\end{equation}
where $\left(  a\right)  _{p}=a\left(  a+1\right)  \left(  a+2\right)
\cdots\left(  a+p-1\right)  $, we expand the local one-loop effective action
(\ref{Wscalar}) around $\left\vert x-y\right\vert =0$:%
\begin{align}
W_{0}\left(  s;q;x,y\right)   &  =-\frac{\tilde{\mu}^{2s}}{2\left(
4\pi\right)  ^{n/2}}\left[  \Gamma\left(  s-\frac{n}{2}\right)  \sum
_{p=0}^{\infty}\frac{\left(  \sqrt{m^{2}+q}\right)  ^{2p+n-2s}}{\left(
1+n/2-s\right)  _{p}p!}\left(  \frac{\left\vert x-y\right\vert }{2}\right)
^{2p}\right. \nonumber\\
&  \left.  +\Gamma\left(  \frac{n}{2}-s\right)  \sum_{p=0}^{\infty}%
\frac{\left(  \sqrt{m^{2}+q}\right)  ^{2p}}{\left(  1-n/2+s\right)  _{p}%
p!}\left(  \frac{\left\vert x-y\right\vert }{2}\right)  ^{2p-n+2s}\right]  .
\end{align}
The negative power terms in the expansion of $W_{0}\left(  s;q;x,y\right)  $
will diverge when $\left\vert x-y\right\vert =0$. That is to say, by such a
procedure, we have extracted the divergent part of $W_{0}\left(
s;q;x,y\right)  $. In order to achieve a finite result, we drop the negative
power term of $\left\vert x-y\right\vert $. Then taking trace gives%
\begin{equation}
W_{0}\left(  s;q\right)  =TrW_{0}\left(  s;q;x,y\right)  =-Vol\frac{\tilde
{\mu}^{2s}}{2\left(  4\pi\right)  ^{n/2}}\left(  m^{2}+q\right)
^{n/2-s}\Gamma\left(  s-\frac{n}{2}\right)  .\label{Ws}%
\end{equation}

In even-dimensional space-times, $s=0$ is a singular point, which is a simple
pole of $W_{0}\left(  s;q\right)  $. In order to remove the divergence coming
from $s=0$, for $2\nu$-dimensional cases, we Laurent expand $W_{0}\left(
s;0\right)  $ around $s=0$.%
\begin{align}
&  W_{0}\left(  s;0\right) \nonumber\\
&  =-Vol\frac{\left(  -1\right)  ^{\nu}}{2\left(  4\pi\right)  ^{\nu}\nu
!}m^{2\nu}\left\{  \frac{1}{s}+\psi\left(  \nu+1\right)  -\ln\frac{m^{2}%
}{\tilde{\mu}^{2}}\right. \nonumber\\
&  +\left.  \sum_{p=2}^{\infty}\left\{  \sum_{\beta=0}^{p}\sum_{\alpha
=0}^{\beta}\frac{\left(  -1\right)  ^{\frac{\alpha+\beta}{2}-1}\left(
2^{\beta-\alpha}-2\right)  B_{\beta-\alpha}\pi^{\beta-\alpha}\Gamma\left(
\nu+1\right)  }{\alpha!\left(  \beta-\alpha\right)  !\left(  p-\beta\right)
!}\left[  \frac{\partial^{\alpha}}{\partial\xi^{\alpha}}\frac{1}{\Gamma\left(
\xi\right)  }\right]  _{\xi=\nu+1}\left(  \ln\frac{\tilde{\mu}^{2}}{m^{2}%
}\right)  ^{p-\beta}\right\}  s^{p-1}\right\}  ,
\end{align}
where $B_{\nu}$ is the Bernoulli number. The finite physical observable does
not contain the parameter $s$, which is embodied in $W_{0}\left(  s;0\right)
$ with $s=0$, i.e., $W_{0}\left(  0;0\right)  $. Take $s=0$. Then, a
regularized unshifted $2\nu$-dimensional one-loop effective action without the
regularization parameter $s$ is the remaining part of the expansion of
$W_{0}\left(  s;0\right)  $ after dropping the divergent negative power term
of $s$, \footnote{In the published version (JHEP06(2010)070), we have used a
misleading usage: we call, e.g., $W\left(  s;0\right)  $ or $W_{s}$, etc., as
regularized one-loop effective action, call $W$, the regularized one-loop
effective action without the regularized parameter $s$ (the remaining part of
$W_{s}$ of taking $s=0$ and dropping the divergent terms of $s$) as
renormalized one-loop effective action, but call the common renormalized
quantities as finite physical quantities.}
\begin{equation}
W_{0}=-Vol\frac{\left(  -1\right)  ^{\nu}}{2\left(  4\pi\right)  ^{\nu}\nu
!}m^{2\nu}\left[  \psi\left(  \nu+1\right)  -\ln\frac{m^{2}}{\tilde{\mu}^{2}%
}\right]  ,
\end{equation}
where $\psi\left(  z\right)  =\Gamma^{\prime}\left(  z\right)  /\Gamma\left(
z\right)  $. The $4$-dimensional case ($\nu=2$) agrees with the result in
Refs. \cite{Avramidi,Vas2}.

In odd-dimensional space-times, $s=0$ is not a singular point, so the one-loop
effective action given by eq. (\ref{Ws}) does not need to drop divergent
terms. The expansion of $W_{0}\left(  s;0\right)  $ reads
\begin{equation}
W_{0}\left(  s;0\right)  =-\frac{Vol}{2\left(  4\pi\right)  ^{\nu+1/2}}%
m^{n}\sum_{p=0}^{\infty}\left[  \sum_{\beta=0}^{p}\frac{\Gamma^{\left(
\beta\right)  }\left(  -\left(  \nu+1/2\right)  \right)  }{\beta!\left(
p-\beta\right)  !}\left(  \ln\frac{\tilde{\mu}^{2}}{m^{2}}\right)  ^{p-\beta
}\right]  s^{p}.
\end{equation}
For $\left(  2\nu+1\right)  $-dimensional cases, from eq. (\ref{Ws}), taking
$s=0$, we arrive at%
\begin{equation}
W_{0}=-Vol\frac{1}{2\left(  4\pi\right)  ^{\nu+1/2}}m^{2\nu+1}\Gamma\left(
-\left(  \nu+1/2\right)  \right)  .
\end{equation}

\subsection{The series solution: the Laplace-type operator with local boundary
conditions}

Exact solutions are rare, so in more general cases, such as interaction cases,
we turn to find perturbation solutions for $W\left(  s;q;x,y\right)  $. When
seeking a perturbation solution, the first thing that we need to do is to find
a proper series expansion for $W\left(  s;q;x,y\right)  $. This is, in
principle, a difficult task. Fortunately, a thorough study on the expansion of
heat kernels has already been made \cite{GS,Grubb}. We can construct a proper
series for one-loop effective actions by starting from the series of heat
kernels, based on the transformation relation between one-loop effective
actions and heat kernels.

In the section, we first consider a special case: the case of a second-order
differential operator of Laplace type with a local boundary condition. In next
section, we discuss the general form of the series expansion of a one-loop
effective action.

In the case of a second-order differential operator of Laplace type with a
local boundary condition, the $n$-dimensional heat kernel corresponding to the
operator $D$ can be expanded as \cite{Vas}%
\begin{equation}
K\left(  t;q;x,y\right)  =K_{0}\left(  t;q;x,y\right)  \sum_{k=0,\frac{1}%
{2},1,\cdots}b_{k}\left(  x,y\right)  t^{k},\label{expansionofKt}%
\end{equation}
where $K_{0}\left(  t;q;x,y\right)  =\left(  4\pi t\right)  ^{-n/2}e^{-\left(
x-y\right)  ^{2}/\left(  4t\right)  -\left(  m^{2}+q\right)  t}$ is the heat
kernel for a $n$-dimensional free massive scalar field and $b_{k}\left(
x,y\right)  $ is the heat kernel coefficient.

The shifted local regularized one-loop effective action can be achieved by
performing the transformation (\ref{ZetaandK}) to the heat kernel $K\left(
t;q;x,y\right)  $ given by eq. (\ref{expansionofKt}):
\begin{align}
W\left(  s;q;x,y\right)   &  =-\frac{\tilde{\mu}^{2s}}{\left(  4\pi\right)
^{n/2}}\sum_{k=0,\frac{1}{2},1,\cdots}b_{k}\left(  x,y\right)  \left(
\frac{2\sqrt{m^{2}+q}}{\left\vert x-y\right\vert }\right)  ^{n/2-k-s}%
\nonumber\\
&  \times K_{n/2-k-s}\left(  \sqrt{m^{2}+q}\left\vert x-y\right\vert \right)
.\label{LocalOLEA}%
\end{align}
$W\left(  s;q;x,y\right)  $ is analytic except for $\left\vert x-y\right\vert
=0$.

In order to achieve the global regularized one-loop effective action, we need
to take trace of eq. (\ref{LocalOLEA}) with $q=0$. The divergence coming from
the singular point $\left\vert x-y\right\vert =0$ of the local one-loop
effective action can be removed by the same procedure used in the case of free
fields. Using eq. (\ref{lim}), we achieve the local regularized one-loop
effective action%
\begin{align}
W\left(  s;q;x,y\right)   &  =-\frac{\tilde{\mu}^{2s}}{2\left(  4\pi\right)
^{n/2}}\sum_{k=0,\frac{1}{2},1,\cdots}b_{k}\left(  x,y\right)  \left[
\Gamma\left(  k-\frac{n}{2}+s\right)  \sum_{p=0}^{\infty}\frac{\left(
m^{2}+q\right)  ^{p-k+n/2-s}\left(  \left\vert x-y\right\vert /2\right)
^{2p}}{p!\left(  1-k+n/2-s\right)  _{p}}\right. \nonumber\\
&  \left.  +\Gamma\left(  -k+\frac{n}{2}-s\right)  \sum_{p=0}^{\infty}%
\frac{\left(  m^{2}+q\right)  ^{p}\left(  \left\vert x-y\right\vert /2\right)
^{2\left(  p-n/2+k+s\right)  }}{p!\left(  1+k-n/2+s\right)  _{p}}\right]  .
\end{align}
In this result, the divergent part of the one-loop effective action has been
extracted. Dropping the negative power term of $\left\vert x-y\right\vert $
and taking trace gives the global result:%
\begin{align}
W\left(  s;q\right)   &  =TrW\left(  s;q;x,y\right) \nonumber\\
&  =-\frac{\tilde{\mu}^{2s}}{2\left(  4\pi\right)  ^{n/2}}\sum_{k=0,\frac
{1}{2},1,\cdots}B_{k}\frac{\Gamma\left(  s-n/2+k\right)  }{\left(
m^{2}+q\right)  ^{k-n/2+s}},
\end{align}
where $B_{k}=trb_{k}\left(  x,y\right)  =\int d^{n}x\sqrt{g}b_{k}\left(
x,x\right)  $.

To obtain a regularized unshifted one-loop effective action, we Laurent expand
$W\left(  s;0\right)  $ with respect to $s$ around $s=0$,%
\begin{align}
&  W\left(  s;0\right) \nonumber\\
&  =-\frac{Vol}{2\left(  4\pi\right)  ^{n/2}}\left\{  \sum
_{\substack{k=0,1/2,1,\cdots\\n/2-k=0,1,2,\cdots}}B_{k}\frac{\left(
-1\right)  ^{n/2-k}}{\left(  n/2-k\right)  !}m^{n-2k}\left\{  \frac{1}{s}%
+\psi\left(  \frac{n}{2}-k+1\right)  -\ln\frac{m^{2}}{\tilde{\mu}^{2}}\right.
\right. \nonumber\\
&  \left.  +\sum_{p=2}^{\infty}\sum_{\beta=0}^{p}\sum_{\alpha=0}^{\beta}%
\frac{\left(  -1\right)  ^{\frac{\alpha+\beta}{2}-1}\left(  2^{\beta-\alpha
}-2\right)  B_{\beta-\alpha}\pi^{\beta-\alpha}\Gamma\left(  n/2-k+1\right)
}{\alpha!\left(  \beta-\alpha\right)  !\left(  p-\beta\right)  !}\left[
\frac{\partial^{\alpha}}{\partial\xi^{\alpha}}\frac{1}{\Gamma\left(
\xi\right)  }\right]  _{\xi=n/2-k+1}\left(  \ln\frac{\tilde{\mu}^{2}}{m^{2}%
}\right)  ^{p-\beta}s^{p-1}\right\} \nonumber\\
&  \left.  +\sum_{\substack{k=0,1/2,1,\cdots\\n/2-k\neq0,1,2,\cdots}}^{\infty
}B_{k}m^{n-2k}\sum_{p=0}^{\infty}\left[  \sum_{\beta=0}^{p}\frac
{\Gamma^{\left(  \beta\right)  }\left(  -n/2+k\right)  }{\beta!\left(
p-\beta\right)  !}\left(  \ln\frac{\tilde{\mu}^{2}}{m^{2}}\right)  ^{p-\beta
}\right]  s^{p}\right\}  .
\end{align}
Take $s=0$ and drop the divergent term:%
\begin{align}
W  &  =\frac{Vol}{2\left(  4\pi\right)  ^{n/2}}\left\{  \sum
_{\substack{k=0,\frac{1}{2},1,\cdots\\\frac{n}{2}-k=0,1,2,\cdots}}B_{k}%
\frac{\left(  -1\right)  ^{n/2-k}}{\left(  n/2-k\right)  !}\frac{1}{m^{2k-n}%
}\left[  -\psi\left(  \frac{n}{2}-k+1\right)  +\ln\frac{m^{2}}{\tilde{\mu}%
^{2}}\right]  \right. \nonumber\\
&  -\left.  \sum_{\substack{k=0,\frac{1}{2},1,\cdots\\\frac{n}{2}%
-k\neq0,1,2,\cdots}}^{\infty}B_{k}\Gamma\left(  k-\frac{n}{2}\right)  \frac
{1}{m^{2k-n}}\right\}  .\label{Wsq}%
\end{align}
For manifolds without boundaries, the half-integer power terms vanish, i.e.,
$B_{m/2}=0$. The result (\ref{Wsq}) with $B_{m/2}=0$ and $n=4$ agrees with the
result given by \cite{Avramidi,Vas2}.

\subsection{The series solution: general cases\label{Wlog}}

In this section, we give a discussion on the general form of series expansion
for local one-loop effective actions, $W\left(  s;q;x,y\right)  $. In order to
achieve a proper series for one-loop effective actions, we start from the
series of heat kernels, based on the transformation relation between one-loop
effective actions and heat kernels.

General form of the heat kernel expansion contains logarithmic terms, which
can be written as \cite{Grubb,Vas}%
\begin{equation}
K\left(  t;q;x,y\right)  =K_{0}\left(  t;q;x,y\right)  \left\{  \sum
_{k=0,\frac{1}{2},1,\cdots}^{N}b_{k}\left(  x,y\right)  t^{k}+\sum
_{k=N+\frac{1}{2}}^{\infty}t^{k}\left[  b_{k}^{\prime}\left(  x,y\right)  \ln
t+b_{k}^{\prime\prime}\left(  x,y\right)  \right]  \right\}
,\label{GexpansionofKt}%
\end{equation}
where $b_{k}\left(  x,y\right)  $, $b_{k}^{\prime}\left(  x,y\right)  $ and
$b_{k}^{\prime\prime}\left(  x,y\right)  $ are heat kernel coefficients.

Starting from the expansion of heat kernels, we can achieve a series of the
local regularized one-loop effective action by performing the transformation
(\ref{ZetaandK}):%
\begin{align}
W\left(  s;q;x,y\right)   &  =-\frac{\tilde{\mu}^{2s}}{\left(  4\pi\right)
^{n/2}}\left\{  \left[  \sum_{k=0,\frac{1}{2},1,\cdots}^{N}b_{k}\left(
x,y\right)  +\sum_{k=N+\frac{1}{2}}^{\infty}b_{k}^{\prime}\left(  x,y\right)
\ln\frac{\left\vert x-y\right\vert }{2\sqrt{m^{2}+q}}+\sum_{k=N+\frac{1}{2}%
}^{\infty}b_{k}^{\prime\prime}\left(  x,y\right)  \right]  \right. \nonumber\\
&  \times\left(  \frac{\left\vert x-y\right\vert }{2\sqrt{m^{2}+q}}\right)
^{k-n/2+s}K_{k-n/2+s}\left(  \sqrt{m^{2}+q}\left\vert x-y\right\vert \right)
\nonumber\\
&  \left.  -\sum_{k=N+\frac{1}{2}}^{\infty}b_{k}^{\prime}\left(  x,y\right)
\left(  \frac{\left\vert x-y\right\vert }{2\sqrt{m^{2}+q}}\right)
^{k-n/2+s}K_{-\left(  k-n/2+s\right)  }^{\left(  1\right)  }\left(
\sqrt{m^{2}+q}\left\vert x-y\right\vert \right)  \right\}  ,\label{GLocalW}%
\end{align}
where $K_{\nu}^{\left(  1\right)  }\left(  z\right)  =$ $\frac{\partial
}{\partial\nu}K_{\nu}\left(  z\right)  $. Based on this series expansion, one
can in principle achieve a perturbation solution of $W\left(  s;q;x,y\right)
$.

In order to achieve a series expansion for the global regularized one-loop
effective action, we take trace of eq. (\ref{GLocalW}). The divergence coming
from the singular point $\left\vert x-y\right\vert =0$ of the local one-loop
effective action can be removed by the same procedure used in the case of free
fields. The series of the shifted global regularized one-loop effective action
then reads%
\begin{align}
&  W\left(  s;q\right)  =trW\left(  s;q;x,y\right) \nonumber\\
&  =-\frac{\tilde{\mu}^{2s}}{2\left(  4\pi\right)  ^{n/2}}\left[
\sum_{k=0,\frac{1}{2},1,\cdots}^{N}B_{k}+\sum_{k=N+\frac{1}{2}}^{\infty}%
B_{k}^{\prime\prime}+\sum_{k=N+\frac{1}{2}}^{\infty}B_{k}^{\prime}\psi\left(
s-\frac{n}{2}+k\right)  \right]  \frac{\Gamma\left(  s-n/2+k\right)  }{\left(
m^{2}+q\right)  ^{k-n/2+s}},\label{WsqSE}%
\end{align}
where the relation%
\begin{align}
K_{\nu}^{\left(  1\right)  }\left(  z\right)   &  =\frac{\pi\csc\left(  \nu
\pi\right)  }{2}\sum_{p=0}^{\infty}\left\{  \left[  \psi\left(  p-\nu
+1\right)  -\frac{\pi}{\tan\left(  \nu\pi\right)  }-\ln\left(  \frac{z}%
{2}\right)  \right]  \frac{1}{\Gamma\left(  p-\nu+1\right)  p!}\left(
\frac{z}{2}\right)  ^{2p-\nu}\right. \nonumber\\
&  +\left.  \left[  \psi\left(  p+\nu+1\right)  +\frac{\pi}{\tan\left(  \nu
\pi\right)  }-\ln\left(  \frac{z}{2}\right)  \right]  \frac{1}{\Gamma\left(
p+\nu+1\right)  p!}\left(  \frac{z}{2}\right)  ^{2p+\nu}\right\}
\end{align}
is used.

To obtain a regularized series expansion without the regularization parameter
$s$, we Laurent expand $W\left(  s;0\right)  $,%
\begin{align}
&  W\left(  s;0\right) \nonumber\\
&  =-\frac{Vol}{2\left(  4\pi\right)  ^{n/2}}\left\{  \left(  \sum
_{\substack{k=0,1/2,1,\cdots\\n/2-k=0,1,2,\cdots}}^{N}B_{k}+\sum
_{\substack{k=N+1/2 \\n/2-k=0,1,2,\cdots}}B_{k}^{\prime\prime}\right)
\frac{\left(  -1\right)  ^{n/2-k}}{\left(  n/2-k\right)  !}m^{n-2k}\right.
\nonumber\\
&  \times\left\{  \frac{1}{s}+\psi\left(  \frac{n}{2}-k+1\right)  -\ln
\frac{m^{2}}{\tilde{\mu}^{2}}\right.  +\sum_{p=2}^{\infty}\left\{  \sum
_{\beta=0}^{p}\sum_{\alpha=0}^{\beta}\frac{\left(  -1\right)  ^{\left(
\alpha+\beta\right)  /2-1}\left(  2^{\beta-\alpha}-2\right)  B_{\beta-\alpha
}\pi^{\beta-\alpha}\Gamma\left(  n/2-k+1\right)  }{\alpha!\left(  \beta
-\alpha\right)  !}\right. \nonumber\\
&  \left.  \times\left.  \left[  \frac{\partial^{\alpha}}{\partial\xi^{\alpha
}}\frac{1}{\Gamma\left(  \xi\right)  }\right]  _{\xi=n/2-k+1}\frac{1}{\left(
p-\beta\right)  !}\left(  \ln\frac{\tilde{\mu}^{2}}{m^{2}}\right)  ^{p-\beta
}\right\}  s^{p-1}\right\} \nonumber\\
&  +\left(  \sum_{\substack{k=0,1/2,1,\cdots\\n/2-k\neq0,1,2,\cdots}}^{N}%
B_{k}+\sum_{\substack{k=N+1/2 \\n/2-k\neq0,1,2,\cdots}}^{\infty}B_{k}%
^{\prime\prime}\right)  m^{n-2k}\sum_{p=0}^{\infty}\left[  \sum_{\beta=0}%
^{p}\frac{\Gamma^{\left(  \beta\right)  }\left(  -n/2+k\right)  }%
{\beta!\left(  p-\beta\right)  !}\left(  \ln\frac{\tilde{\mu}^{2}}{m^{2}%
}\right)  ^{p-\beta}\right]  s^{p}\nonumber
\end{align}%
\begin{align}
&  +\sum_{\substack{k=N+1/2 \\n/2-k=0,1,2,\cdots}}B_{k}^{\prime}m^{n-2k}%
\frac{\left(  -1\right)  ^{n/2-k}}{\left(  n/2-k\right)  !}\nonumber\\
&  \times\left\{  -\frac{1}{s^{2}}-\frac{1}{s}\ln\frac{\tilde{\mu}^{2}}{m^{2}%
}+\left[  \frac{1}{2}\psi^{2}\left(  \frac{n}{2}-k+1\right)  -\frac{1}{2}%
\psi^{\left(  1\right)  }\left(  \frac{n}{2}-k+1\right)  -\frac{1}{2}\left(
\ln\frac{\tilde{\mu}^{2}}{m^{2}}\right)  ^{2}+\frac{\pi^{2}}{6}\right]
\right. \nonumber\\
&  +\sum_{p=3}^{\infty}\left\{  \sum_{\beta=0}^{p}\frac{\left(  \beta
-1\right)  }{\left(  p-\beta\right)  !}\sum_{\alpha=0}^{\beta}\frac{\left(
-1\right)  ^{\left(  \alpha+\beta\right)  /2-1}\left(  2^{\beta-\alpha
}-2\right)  B_{\beta-\alpha}\pi^{\beta-\alpha}\Gamma\left(  n/2-k+1\right)
}{\alpha!\left(  \beta-\alpha\right)  !}\right. \nonumber\\
&  \left.  \times\left.  \left[  \frac{\partial^{\alpha}}{\partial\xi^{\alpha
}}\frac{1}{\Gamma\left(  \xi\right)  }\right]  _{\xi=n/2-k+1}\left(  \ln
\frac{\tilde{\mu}^{2}}{m^{2}}\right)  ^{p-\beta}\right\}  s^{p-2}\right\}
\nonumber\\
&  \left.  +\sum_{\substack{k=N+1/2 \\n/2-k\neq0,1,2,\cdots}}^{\infty}%
B_{k}^{\prime}m^{n-2k}\sum_{p=0}^{\infty}\left[  \sum_{\beta=0}^{p}%
\frac{\Gamma^{\left(  \beta+1\right)  }\left(  -n/2+k\right)  }{\beta!\left(
p-\beta\right)  !}\left(  \ln\frac{\tilde{\mu}^{2}}{m^{2}}\right)  ^{p-\beta
}\right]  s^{p}\right\}  .
\end{align}
and then take $s=0$ and drop the divergent terms,%
\begin{align}
W  &  =-\frac{1}{2\left(  4\pi\right)  ^{n/2}}\left\{  \left\{  \sum
_{\substack{k=0,\frac{1}{2},1,\cdots\\\frac{n}{2}-k\neq0,1,2,\cdots}}^{N}%
B_{k}+\sum_{\substack{k=N+\frac{1}{2} \\\frac{n}{2}-k\neq0,1,2,\cdots
}}^{\infty}\left[  B_{k}^{\prime\prime}+B_{k}^{\prime}\psi\left(  k-\frac
{n}{2}\right)  \right]  \right\}  \Gamma\left(  k-\frac{n}{2}\right)
m^{n-2k}\right. \nonumber\\
&  +\left(  \sum_{\substack{k=0,\frac{1}{2},1,\cdots\\\frac{n}{2}%
-k=0,1,2,\cdots}}^{N}B_{k}+\sum_{\substack{k=N+\frac{1}{2} \\\frac{n}%
{2}-k=0,1,2,\cdots}}B_{k}^{\prime\prime}\right)  \frac{\left(  -1\right)
^{-n/2+k}}{\left(  n/2-k\right)  !}m^{n-2k}\left[  \psi\left(  1-k+\frac{n}%
{2}\right)  -\ln\frac{m^{2}}{\tilde{\mu}^{2}}\right] \nonumber\\
&  +\left.  \sum_{\substack{k=N+\frac{1}{2} \\\frac{n}{2}-k=0,1,2,\cdots
}}B_{k}^{\prime}\frac{\left(  -1\right)  ^{-n/2+k}}{\left(  n/2-k\right)
!}m^{n-2k}\left[  \frac{1}{2}\psi^{2}\left(  1-k+\frac{n}{2}\right)  -\frac
{1}{2}\psi^{\prime}\left(  1-k+\frac{n}{2}\right)  +\frac{\pi^{2}}{6}-\frac
{1}{2}\left(  \ln\frac{m^{2}}{\tilde{\mu}^{2}}\right)  ^{2}\right]  \right\}
,
\end{align}
where $\psi^{\prime}\left(  z\right)  =\frac{d}{dz}\psi\left(  z\right)  $.

\section{Solutions of local and global vacuum energies: Renormalization
\label{vacuumenergy}}

\subsection{The free-field solution}

For a free massive scalar field in $\mathbb{R}^{n}$, $D_{0}=-\nabla^{2}+m^{2}
$. The shifted local vacuum energy can be solved from eq. (\ref{EqofveDeq}):%
\begin{equation}
E_{0}\left(  \epsilon;q;x,y\right)  =\frac{\tilde{\mu}^{2\epsilon}}{\left(
4\pi\right)  ^{n/2}\Gamma\left(  -1/2+\epsilon\right)  }\left(  \frac
{2\sqrt{m^{2}+q}}{\left\vert x-y\right\vert }\right)  ^{\left(  n+1\right)
/2-\epsilon}K_{-\left(  n+1\right)  /2+\epsilon}\left(  \sqrt{m^{2}%
+q}\left\vert x-y\right\vert \right)  .
\end{equation}

The shifted local vacuum energy $E_{0}\left(  \epsilon;q;x,y\right)  $ has a
singular point at $\left\vert x-y\right\vert =0$, corresponding to the
divergence in the global vacuum energy which is the trace of the local one. In
order to extract the divergence, we expand $E_{0}\left(  \epsilon
;q;x,y\right)  $ around the singularity $\left\vert x-y\right\vert =0$,%
\begin{align}
E_{0}\left(  \epsilon;q;x,y\right)   &  =\frac{\tilde{\mu}^{2\epsilon}%
}{2\left(  4\pi\right)  ^{n/2}\Gamma\left(  -1/2+\epsilon\right)  }\frac{\pi
}{\sin\left(  \left(  -\left(  n+1\right)  /2+\epsilon\right)  \pi\right)
}\nonumber\\
&  \times\sum_{p=0}^{\infty}\frac{1}{p!}\left[  \frac{\left(  m^{2}+q\right)
^{\left(  n+1\right)  /2+p-\epsilon}\left(  \left\vert x-y\right\vert
/2\right)  ^{2p}}{\Gamma\left(  1+\left(  n+1\right)  /2-\epsilon+p\right)
}-\frac{\left(  m^{2}+q\right)  ^{p}\left(  \left\vert x-y\right\vert
/2\right)  ^{2p-n-1+2\epsilon}}{\Gamma\left(  1-\left(  n+1\right)
/2+\epsilon+p\right)  }\right]  .
\end{align}
Taking trace and dropping the divergent negative power term gives%
\begin{equation}
E_{0}\left(  \epsilon;q\right)  =Vol\frac{\tilde{\mu}^{2\epsilon}}{2\left(
4\pi\right)  ^{n/2}}\frac{\Gamma\left(  -\left(  n+1\right)  /2+\epsilon
\right)  }{\Gamma\left(  -1/2+\epsilon\right)  }\left(  m^{2}+q\right)
^{\left(  n+1\right)  /2-\epsilon}.\label{E0efree}%
\end{equation}
In odd-dimensional space-times, $\epsilon=0$ is a singular point of
$E_{0}\left(  \epsilon;q\right)  $. To remove the divergence, we Laurent
expand $E_{0}\left(  \epsilon;0\right)  $ with respect to $\epsilon$ around
$\epsilon=0$,%
\begin{align}
&  E_{0}\left(  \epsilon\right) \nonumber\\
&  =-Vol\frac{m^{2\nu}}{2\left(  4\pi\right)  ^{\nu}}\frac{\left(  -1\right)
^{\nu}}{\nu!}\left\{  \frac{1}{\epsilon}+\left[  \ln\frac{4\tilde{\mu}^{2}%
}{m^{2}}+\psi\left(  \nu+1\right)  +\gamma_{E}-2\right]  \right. \nonumber\\
&  +\frac{\epsilon}{12}\left[  12\left(  \ln\frac{4\tilde{\mu}^{2}}{m^{2}%
}+\gamma_{E}-2\right)  \psi\left(  \nu+1\right)  +6\left(  \ln\frac
{4\tilde{\mu}^{2}}{m^{2}}+\gamma_{E}-2\right)  ^{2}+6\psi^{2}\left(
\nu+1\right)  -6\psi^{\left(  1\right)  }\left(  \nu+1\right)  -\pi
^{2}-24\right] \nonumber\\
&  +\frac{\epsilon^{2}}{12}\left\{  6\left(  \ln\frac{4\tilde{\mu}^{2}}{m^{2}%
}+\gamma_{E}-2\right)  \psi^{2}\left(  \nu+1\right)  +\psi\left(
\nu+1\right)  \left[  6\ln\frac{4\tilde{\mu}^{2}}{m^{2}}\left(  \ln
\frac{4\tilde{\mu}^{2}}{m^{2}}+2\gamma_{E}-4\right)  \right.  \right.
\nonumber\\
&  \left.  -6\psi^{\left(  1\right)  }\left(  \nu+1\right)  +6\gamma
_{E}\left(  \gamma_{E}-4\right)  -\pi^{2}\right]  -6\left(  \ln\frac
{4\tilde{\mu}^{2}}{m^{2}}+\gamma_{E}-2\right)  \psi^{\left(  1\right)
}\left(  \nu+1\right) \nonumber\\
&  +\ln\frac{4\tilde{\mu}^{2}}{m^{2}}\left[  2\ln\frac{4\tilde{\mu}^{2}}%
{m^{2}}\left(  \ln\frac{4\tilde{\mu}^{2}}{m^{2}}+3\gamma_{E}-6\right)
+6\gamma_{E}\left(  \gamma_{E}-4\right)  -\pi^{2}\right] \nonumber\\
&  \left.  \left.  +2\psi^{3}\left(  \nu+1\right)  +2\psi^{2}\left(
\nu+1\right)  +28\zeta\left(  3\right)  +2\pi^{2}+\gamma_{E}\left[  2\left(
\gamma_{E}-6\right)  \gamma_{E}-\pi^{2}\right]  \right\}  +\cdots\right\}  .
\end{align}
A regularized unshifted $\left(  2\nu-1\right)  $-dimensional vacuum energy
without the regularization parameter $s$ can be obtained by taking
$\epsilon=0$ and dropping the divergent negative power term of $\epsilon$:%
\begin{equation}
E_{0}=Vol\frac{\left(  -1\right)  ^{\nu}}{2\left(  4\pi\right)  ^{\nu}\nu
!}m^{2\nu}\left[  2-\gamma_{E}-\psi\left(  \nu+1\right)  +\ln\frac{m^{2}%
}{4\tilde{\mu}^{2}}\right]  ,
\end{equation}
where $\gamma_{E}$ is the Euler constant.

In even-dimensional space-times, $\epsilon=0$ is not a singular point, so the
$2\nu$-dimensional vacuum energy can be achieved directly by setting
$\epsilon=0$ in eq. (\ref{E0efree}) without dropping divergent terms. The
expansion of $E_{0}\left(  \epsilon\right)  $ is%
\begin{align}
E_{0}\left(  \epsilon\right)   &  =-Vol\frac{m^{2\nu+1}}{2\left(  4\pi\right)
^{\nu+1/2}}\frac{\Gamma\left(  -1/2\right)  \Gamma\left(  -\left(
\nu+1/2\right)  +\epsilon\right)  }{\Gamma\left(  -1/2+\epsilon\right)
}\left(  \frac{\tilde{\mu}^{2}}{m^{2}}\right)  ^{\epsilon}\nonumber\\
&  =-Vol\frac{m^{2\nu+1}}{2\left(  4\pi\right)  ^{\nu+1/2}}\Gamma\left(
-\left(  \nu+1/2\right)  \right)  \left\{  1+\epsilon\left[  \ln\frac
{4\tilde{\mu}^{2}}{m^{2}}+H_{-\left(  \nu+3/2\right)  }-2\right]  \right.
\nonumber\\
&  +\frac{\epsilon^{2}}{4}\left[  2\psi\left(  -\left(  \nu+1/2\right)
\right)  \left(  2\ln\frac{4\tilde{\mu}^{2}}{m^{2}}+H_{-\left(  \nu
+3/2\right)  }+\gamma_{E}-4\right)  +2\ln\frac{4\tilde{\mu}^{2}}{m^{2}}\left(
\ln\frac{4\tilde{\mu}^{2}}{m^{2}}+2\gamma_{E}-4\right)  \right. \nonumber\\
&  +\left.  \left.  2\psi^{\left(  1\right)  }\left(  -\left(  \nu+1/2\right)
\right)  +2\gamma_{E}\left(  \gamma_{E}-4\right)  -\pi^{2}\right]
+\cdots\right\}  .
\end{align}
Taking $\epsilon=0$ gives
\begin{equation}
E_{0}=-Vol\frac{1}{2\left(  2\sqrt{\pi}\right)  ^{2\nu+1}}\Gamma\left(
-\left(  \nu+1/2\right)  \right)  m^{2\nu+1}.
\end{equation}

\subsection{The series solution: the Laplace-type operator with local boundary
conditions}

To find a perturbation solution for the vacuum energy, we need to first
construct a proper series for $E_{0}\left(  \epsilon;q;x,y\right)  $. In this
section, we first consider the case of a second-order differential operator of
Laplace type $D$ with a local boundary condition.

The series expansion of a $n$-dimensional vacuum energy can be obtained by
performing the transformation (\ref{ZetaandK}) to eq. (\ref{expansionofKt}):%
\begin{align}
&  E_{0}\left(  \epsilon;q;x,y\right)  =\frac{\tilde{\mu}^{2\epsilon}}{\left(
4\pi\right)  ^{n/2}\Gamma\left(  -1/2+\epsilon\right)  }\nonumber\\
&  \times\sum_{k=0,\frac{1}{2},1,\cdots}b_{k}\left(  x,y\right)  \left(
\frac{\left\vert x-y\right\vert }{2\sqrt{m^{2}+q}}\right)  ^{k-\left(
n+1\right)  /2+\epsilon}K_{-\left(  n+1\right)  /2+k+\epsilon}\left(
\sqrt{m^{2}+q}\left\vert x-y\right\vert \right)  .
\end{align}
Taking trace and dropping the divergent negative power term gives%
\begin{equation}
E_{0}\left(  \epsilon;q\right)  =\frac{\tilde{\mu}^{2\epsilon}}{2\left(
4\pi\right)  ^{n/2}\Gamma\left(  -1/2+\epsilon\right)  }\sum_{k=0,\frac{1}%
{2},1,\cdots}B_{k}\Gamma\left(  -\frac{n+1}{2}+k+\epsilon\right)  \left(
m^{2}+q\right)  ^{\left(  n+1\right)  /2-k-\epsilon}.
\end{equation}
To achieve an unshifted regularized vacuum energy, we Laurent expand
$E_{0}\left(  \epsilon;0\right)  $,%
\begin{align}
E_{0}\left(  \epsilon\right)   &  =-\frac{Vol}{2\left(  4\pi\right)  ^{\left(
n+1\right)  /2}}\left\{  \frac{1}{\epsilon}\sum_{\substack{k=0,\frac{1}%
{2},1,\cdots\\\left(  n+1\right)  /2-k=0,1,2,\cdots}}B_{k}m^{\left(
n+1\right)  -2k}\frac{\left(  -1\right)  ^{\left(  n+1\right)  /2-k}}{\left(
\left(  n+1\right)  /2-k\right)  !}\right. \nonumber\\
&  +\left\{  \sum_{\substack{k=0,\frac{1}{2},1,\cdots\\\left(  n+1\right)
/2-k=0,1,2,\cdots}}B_{k}m^{\left(  n+1\right)  -2k}\frac{\left(  -1\right)
^{\left(  n+1\right)  /2-k}}{\left(  \left(  n+1\right)  /2-k\right)
!}\left[  \ln\frac{4\tilde{\mu}^{2}}{m^{2}}+\psi\left(  \frac{n+3}%
{2}-k\right)  +\gamma_{E}-2\right]  \right. \nonumber\\
&  \left.  +\sum_{\substack{k=0,\frac{1}{2},1,\cdots\\\left(  n+1\right)
/2-k\neq0,1,2,\cdots}}^{\infty}B_{k}m^{\left(  n+1\right)  -2k}\Gamma\left(
-\frac{n+1}{2}+k\right)  \right\} \nonumber\\
&  +\epsilon\left\{  \frac{1}{12}\sum_{\substack{k=0,\frac{1}{2}%
,1,\cdots\\\left(  n+1\right)  /2-k=0,1,2,\cdots}}B_{k}m^{\left(  n+1\right)
-2k}\frac{\left(  -1\right)  ^{\left(  n+1\right)  /2-k}}{\left(  \left(
n+1\right)  /2-k\right)  !}\left[  12\left(  \ln\frac{4\tilde{\mu}^{2}}{m^{2}%
}+\gamma_{E}-2\right)  \psi\left(  \frac{n+3}{2}-k\right)  \right.  \right.
\nonumber\\
&  \left.  +6\left(  \ln\frac{4\tilde{\mu}^{2}}{m^{2}}+\gamma_{E}-2\right)
^{2}+6\psi^{2}\left(  \frac{n+3}{2}-k\right)  -6\psi^{\left(  1\right)
}\left(  \frac{n+3}{2}-k\right)  -\pi^{2}-24\right] \nonumber\\
&  \left.  \left.  +\sum_{\substack{k=0,\frac{1}{2},1,\cdots\\\left(
n+1\right)  /2-k\neq0,1,2,\cdots}}^{\infty}B_{k}m^{\left(  n+1\right)
-2k}\Gamma\left(  -\frac{n+1}{2}+k\right)  \left[  \ln\frac{4\tilde{\mu}^{2}%
}{m^{2}}+H_{-\left(  n+3\right)  /2+k}-2\right]  \right\}  +\cdots\right\}  .
\end{align}
Taking $\epsilon=0$ and dropping the divergent negative power term gives%
\begin{align}
E_{0}  &  =\frac{1}{2\left(  4\pi\right)  ^{\left(  n+1\right)  /2}}\left\{
\sum_{\substack{k=0,\frac{1}{2},1,\cdots\\\frac{n+1}{2}-k=0,1,2,\cdots}%
}B_{k}\frac{\left(  -1\right)  ^{\left(  n+1\right)  /2-k}m^{n+1-2k}}{\left[
\left(  n+1\right)  /2-k\right]  !}\left[  2-\gamma_{E}-\psi\left(  \frac
{n+3}{2}-k\right)  +\ln\frac{m^{2}}{4\tilde{\mu}^{2}}\right]  \right.
\nonumber\\
&  -\left.  \sum_{\substack{k=0,\frac{1}{2},1,\cdots\\\frac{n+1}{2}%
-k\neq0,1,2,\cdots}}^{\infty}B_{k}\Gamma\left(  k-\frac{n+1}{2}\right)
m^{n+1-2k}\right\}  .
\end{align}

\subsection{The series solution: general cases}

To construct the general expansion for vacuum energies, we start from the
general form of the expansion of heat kernels, eq. (\ref{GexpansionofKt}). By
eqs. (\ref{veo}) and (\ref{ZetaandK}), we arrive at%
\begin{align}
&  E_{0}\left(  \epsilon;q;x,y\right) \nonumber\\
&  =\frac{\tilde{\mu}^{2\epsilon}}{\left(  4\pi\right)  ^{n/2}\Gamma\left(
-1/2+\epsilon\right)  }\left\{  \left[  \sum_{k=0,\frac{1}{2},1,\cdots}%
^{N}b_{k}\left(  x,y\right)  +\sum_{k=N+\frac{1}{2}}^{\infty}b_{k}^{\prime
}\left(  x,y\right)  \ln\frac{\left\vert x-y\right\vert }{2\sqrt{m^{2}+q}%
}+\sum_{k=N+\frac{1}{2}}^{\infty}b_{k}^{\prime\prime}\left(  x,y\right)
\right]  \right. \nonumber\\
&  \times\left(  \frac{\left\vert x-y\right\vert }{2\sqrt{m^{2}+q}}\right)
^{k-\left(  n+1\right)  /2+\epsilon}K_{k-\left(  n+1\right)  /2+\epsilon
}\left(  \sqrt{m^{2}+q}\left\vert x-y\right\vert \right) \nonumber\\
&  \left.  -\sum_{k=N+\frac{1}{2}}^{\infty}b_{k}^{\prime}\left(  x,y\right)
\left(  \frac{\left\vert x-y\right\vert }{2\sqrt{m^{2}+q}}\right)  ^{k-\left(
n+1\right)  /2+\epsilon}K_{-k+\left(  n+1\right)  /2-\epsilon}^{\left(
1\right)  }\left(  \sqrt{m^{2}+q}\left\vert x-y\right\vert \right)  \right\}
.
\end{align}

Taking trace and dropping the divergent term gives the shifted global vacuum
energy:%
\begin{align}
E_{0}\left(  \epsilon;q\right)   &  =TrE_{0}\left(  \epsilon;q;x,y\right)
\nonumber\\
&  =\frac{\tilde{\mu}^{2\varepsilon}}{2\left(  4\pi\right)  ^{n/2}%
\Gamma\left(  -1/2+\epsilon\right)  }\nonumber\\
&  \times\left[  \sum_{k=0,\frac{1}{2},1,\cdots}^{N}B_{k}+\sum_{k=N+\frac
{1}{2}}^{\infty}B_{k}^{\prime\prime}+\sum_{k=N+\frac{1}{2}}^{\infty}%
B_{k}^{\prime}\psi\left(  k-\frac{n+1}{2}+\epsilon\right)  \right]
\frac{\Gamma\left(  k-\left(  n+1\right)  /2+\epsilon\right)  }{\left(
m^{2}+q\right)  ^{k-\left(  n+1\right)  /2+\epsilon}}.
\end{align}

In order to extract the divergence corresponding to $\epsilon=0$, we Laurent
expand $E_{0}\left(  \epsilon;0\right)  $,%
\begin{align*}
&  E_{0}\left(  \epsilon\right) \\
&  =-Vol\frac{1}{2\left(  4\pi\right)  ^{\left(  n+1\right)  /2}}\left\{
-\frac{1}{\epsilon^{2}}\sum_{\substack{k=N+1/2 \\\left(  n+1\right)
/2-k=0,1,2,\cdots}}B_{k}^{\prime}m^{\left(  n+1\right)  -2k}\frac{\left(
-1\right)  ^{-\left(  n+1\right)  /2+k}}{\left(  \left(  n+1\right)
/2-k\right)  !}\right. \\
&  +\frac{1}{\epsilon}\left\{  \left[  \sum_{\substack{k=0,1/2,1,\cdots
\\\left(  n+1\right)  /2-k=0,1,2,\cdots}}^{N}B_{k}+\sum_{\substack{k=N+1/2
\\\left(  n+1\right)  /2-k=0,1,2,\cdots}}B_{k}^{\prime\prime}\right]
m^{\left(  n+1\right)  -2k}\frac{\left(  -1\right)  ^{\left(  n+1\right)
/2-k}}{\left(  \left(  n+1\right)  /2-k\right)  !}\right. \\
&  \left.  -\sum_{\substack{k=N+1/2 \\\left(  n+1\right)  /2-k=0,1,2,\cdots
}}B_{k}^{\prime}m^{\left(  n+1\right)  -2k}\frac{\left(  -1\right)  ^{-\left(
n+1\right)  /2+k}}{\left(  \left(  n+1\right)  /2-k\right)  !}\left(  \ln
\frac{4\tilde{\mu}^{2}}{m^{2}}+\gamma_{E}-2\right)  \right\} \\
&  +\left\{  \left[  \sum_{\substack{k=0,1/2,1,\cdots\\\left(  n+1\right)
/2-k=0,1,2,\cdots}}^{N}B_{k}+\sum_{\substack{k=N+1/2 \\\left(  n+1\right)
/2-k=0,1,2,\cdots}}B_{k}^{\prime\prime}\right]  \right. \\
&  \times m^{\left(  n+1\right)  -2k}\frac{\left(  -1\right)  ^{\left(
n+1\right)  /2-k}}{\left(  \left(  n+1\right)  /2-k\right)  !}\left[  \ln
\frac{4\tilde{\mu}^{2}}{m^{2}}+\psi\left(  \frac{n+3}{2}-k\right)  +\gamma
_{E}-2\right]
\end{align*}%
\begin{align}
&  +\left[  \sum_{\substack{k=0,1/2,1,\cdots\\\left(  n+1\right)
/2-k\neq0,1,2,\cdots}}^{N}B_{k}+\sum_{\substack{k=N+1/2 \\\left(  n+1\right)
/2-k\neq0,1,2,\cdots}}^{\infty}B_{k}^{\prime\prime}\right]  m^{\left(
n+1\right)  -2k}\Gamma\left(  -\frac{n+1}{2}+k\right) \nonumber\\
&  -\frac{1}{2}\sum_{\substack{k=N+1/2 \\\left(  n+1\right)  /2-k=0,1,2,\cdots
}}B_{k}^{\prime}m^{\left(  n+1\right)  -2k}\frac{\left(  -1\right)  ^{-\left(
n+1\right)  /2+k}}{\left(  \left(  n+1\right)  /2-k\right)  !}\nonumber\\
&  \times\left[  4\ln\frac{2\tilde{\mu}}{m}\left(  \ln\frac{2\tilde{\mu}}%
{m}+\gamma_{E}-2\right)  -\psi^{2}\left(  \frac{n+3}{2}-k\right)
+\psi^{\left(  1\right)  }\left(  \frac{n+3}{2}-k\right)  -\frac{5\pi^{2}}%
{6}+\left(  \gamma_{E}-4\right)  \gamma_{E}\right] \nonumber\\
&  \left.  +\sum_{\substack{k=N+1/2 \\\left(  n+1\right)  /2-k\neq
0,1,2,\cdots}}^{\infty}B_{k}^{\prime}m^{\left(  n+1\right)  -2k}\Gamma\left(
-\frac{n+1}{2}+k\right)  \psi\left(  -\frac{n+1}{2}+k\right)  \right\}
\nonumber\\
&  +\epsilon\left\{  \left[  \sum_{\substack{k=0,1/2,1,\cdots\\\left(
n+1\right)  /2-k=0,1,2,\cdots}}^{N}B_{k}+\sum_{\substack{k=N+1/2 \\\left(
n+1\right)  /2-k=0,1,2,\cdots}}B_{k}^{\prime\prime}\right]  m^{\left(
n+1\right)  -2k}\frac{\left(  -1\right)  ^{\left(  n+1\right)  /2-k}}{\left(
\left(  n+1\right)  /2-k\right)  !}\right. \nonumber\\
&  \times\left\{  \left[  \left(  \ln\frac{4\tilde{\mu}^{2}}{m^{2}}+\gamma
_{E}-2\right)  \psi\left(  \frac{n+3}{2}-k\right)  +\frac{1}{2}\left(
\ln\frac{4\tilde{\mu}^{2}}{m^{2}}+\gamma_{E}-2\right)  ^{2}\right.  \right.
\nonumber\\
&  \left.  +\left.  \frac{1}{2}\psi^{2}\left(  \frac{n+3}{2}-k\right)
-\frac{1}{2}\psi^{\left(  1\right)  }\left(  \frac{n+3}{2}-k\right)
-\frac{\pi^{2}}{12}-2\right]  \right\} \nonumber\\
&  +\left[  \sum_{\substack{k=0,1/2,1,\cdots\\\left(  n+1\right)
/2-k\neq0,1,2,\cdots}}^{N}B_{k}+\sum_{\substack{k=N+1/2 \\\left(  n+1\right)
/2-k\neq0,1,2,\cdots}}^{\infty}B_{k}^{\prime\prime}\right]  m^{\left(
n+1\right)  -2k}\Gamma\left(  -\frac{n+1}{2}+k\right) \nonumber\\
&  \times\left[  \ln\frac{4\tilde{\mu}^{2}}{m^{2}}+H_{-\left(  n+3\right)
/2+k}-2\right]  -\frac{1}{12}\sum_{\substack{k=N+1/2 \\\left(  n+1\right)
/2-k=0,1,2,\cdots}}B_{k}^{\prime}m^{\left(  n+1\right)  -2k}\frac{\left(
-1\right)  ^{-\left(  n+1\right)  /2+k}}{\left(  \left(  n+1\right)
/2-k\right)  !}\nonumber\\
&  \times\left\{  -6\left(  \ln\frac{4\tilde{\mu}^{2}}{m^{2}}+\gamma
_{E}-2\right)  \psi^{2}\left(  \frac{n+3}{2}-k\right)  +6\left(  \ln
\frac{4\tilde{\mu}^{2}}{m^{2}}+\gamma_{E}-2\right)  \psi^{\left(  1\right)
}\left(  \frac{n+3}{2}-k\right)  \right. \nonumber\\
&  +\ln\frac{4\tilde{\mu}^{2}}{m^{2}}\left[  2\ln\frac{4\tilde{\mu}^{2}}%
{m^{2}}\left(  \ln\frac{4\tilde{\mu}^{2}}{m^{2}}+3\gamma_{E}-6\right)
+6\gamma_{E}\left(  \gamma_{E}-4\right)  -5\pi^{2}\right] \nonumber\\
&  \left.  +28\zeta\left(  3\right)  -4\psi^{3}\left(  \frac{n+3}{2}-k\right)
-4\left[  \pi^{2}-3\psi^{\left(  1\right)  }\left(  \frac{n+3}{2}-k\right)
\right]  \psi\left(  \frac{n+3}{2}-k\right)  -4\psi^{\left(  2\right)
}\left(  \frac{n+3}{2}-k\right)  \right\} \nonumber\\
&  +10\pi^{2}+\gamma_{E}\left[  2\left(  \gamma_{E}-6\right)  \gamma_{E}%
-5\pi^{2}\right] \nonumber\\
&  +\sum_{\substack{k=N+1/2 \\\left(  n+1\right)  /2-k\neq0,1,2,\cdots
}}^{\infty}B_{k}^{\prime}m^{\left(  n+1\right)  -2k}\Gamma\left(  -\frac
{n+1}{2}+k\right) \nonumber\\
\times &  \left.  \left.  \left[  \psi\left(  -\frac{n+1}{2}+k\right)  \left(
\ln\frac{4\tilde{\mu}^{2}}{m^{2}}+\gamma_{E}-2+\psi\left(  -\frac{n+1}%
{2}+k\right)  \right)  +\psi^{\left(  1\right)  }\left(  -\frac{n+1}%
{2}+k\right)  \right]  \right\}  +\cdots\right\}  .
\end{align}
Taking $\epsilon=0$ and dropping the divergent term gives the series expansion
of the unshifted regularized vacuum energy without the regularization
parameter $\epsilon$,%
\begin{align}
E_{0} &  =-\frac{1}{2\left(  4\pi\right)  ^{\left(  n+1\right)  /2}}\left\{
\left\{  \sum_{\substack{k=0,\frac{1}{2},1,\cdots\\\frac{n+1}{2}%
-k\neq0,1,2,\cdots}}^{N}B_{k}+\sum_{\substack{k=N+\frac{1}{2} \\\frac{n+1}%
{2}-k\neq0,1,2,\cdots}}^{\infty}\left[  B_{k}^{\prime\prime}+B_{k}^{\prime
}\psi\left(  k-\frac{n+1}{2}\right)  \right]  \right\}  \right. \nonumber\\
&  \times\frac{\Gamma\left(  k-\left(  n+1\right)  /2\right)  }{m^{2k-n-1}%
}+\left[  \sum_{\substack{k=0,\frac{1}{2},1,\cdots\\\frac{n+1}{2}%
-k=0,1,2,\cdots}}^{N}B_{k}+\sum_{\substack{k=N+\frac{1}{2} \\\frac{n+1}%
{2}-k=0,1,2,\cdots}}B_{k}^{\prime\prime}\right]  \frac{\left(  -1\right)
^{-\left(  n+1\right)  /2+k}}{\left(  \left(  n+1\right)  /2-k\right)
!}m^{n+1-2k}\nonumber\\
&  \times\left[  H_{\left(  n+1\right)  /2-k}-2-\ln\frac{m^{2}}{4\tilde{\mu
}^{2}}\right]  +\sum_{\substack{k=N+\frac{1}{2} \\\frac{n+1}{2}-k=0,1,2,\cdots
}}^{\infty}B_{k}^{\prime}\frac{\left(  -1\right)  ^{-\left(  n+1\right)
/2+k}}{\left(  \left(  n+1\right)  /2-k\right)  !}m^{n+1-2k}\nonumber\\
&  \times\left.  \left[  \frac{1}{2}\psi^{2}\left(  \frac{n+3}{2}-k\right)
-\frac{1}{2}\left(  \ln\frac{m^{2}}{4\tilde{\mu}^{2}}-\gamma_{E}+2\right)
^{2}-\frac{1}{2}\psi^{\left(  1\right)  }\left(  \frac{n+3}{2}-k\right)
+\frac{5\pi^{2}}{12}+2\right]  \right\}  .
\end{align}

\section{Solutions of local and global spectral counting functions:
Renormalization \label{spectralcountingfunction}}

\subsection{The free-field solution}

We now solve the spectral counting function from eq. (\ref{dieqofN}) for a
free massive scalar field in $\mathbb{R}^{n}$; in this case, $D_{0}%
=-\nabla^{2}+m^{2}$.

The solution of eq. (\ref{dieqofN}) for $D_{0}$ reads%
\begin{equation}
N_{0}\left(  \lambda;x,y\right)  =\left(  \frac{\sqrt{\lambda-m^{2}}}%
{2\pi\left\vert x-y\right\vert }\right)  ^{n/2}J_{n/2}\left(  \left\vert
x-y\right\vert \sqrt{\lambda-m^{2}}\right)  ,
\end{equation}
where $J_{k}\left(  z\right)  $ is the Bessel function of the first kind.

The spectral counting function can be obtained by taking trace of
$N_{0}\left(  \lambda;x,y\right)  $:%
\begin{equation}
N_{0}\left(  \lambda\right)  =Vol\frac{\left(  \lambda-m^{2}\right)  ^{n/2}%
}{\left(  4\pi\right)  ^{n/2}\Gamma\left(  1+n/2\right)  }.
\end{equation}
The case of $m=0$ and $n=2$ recovers Weyl's famous result \cite{Kac}.

\subsection{The series solution: the Laplace-type operator with local boundary
conditions}

In order to seek an approximation solution for the counting function, we need
to construct a proper series for $N\left(  \lambda;x,y\right)  $. When the
problem is the Laplace-type operator with local boundary conditions, we can
start from the expansion of the heat kernel, eq. (\ref{expansionofKt}).

At the first sight, it seems that one can achieve the expansion of $N\left(
\lambda;x,y\right)  $ by performing the integral transformation (\ref{KNq}) to
eq. (\ref{expansionofKt}) with $q=0$ directly. However, the series
(\ref{expansionofKt}) is not uniformly convergent, so the integral
transformation cannot be applied term by term, i.e., the order of integral and
summation cannot be exchanged. As a result, when performing the integral
transformation term by term, some of the terms will diverge. Concretely, when
applying the transformation (\ref{KNq}) to each term of eq.
(\ref{expansionofKt}), one encounters the integral%
\begin{equation}
\frac{1}{2\pi i}\int_{c-i\infty}^{c+i\infty}\frac{e^{\lambda t}}{t}\left(
4\pi t\right)  ^{-n/2}e^{-\left(  x-y\right)  ^{2}/\left(  4t\right)  -m^{2}%
t}t^{k}dt;\label{int}%
\end{equation}
when $k\geq n/2+1$, the integral diverges. To make sense of these divergent
integrals, we need a renormalization procedure for removing the divergence.

When $k<n/2+1$, the integral which equals the Bessel function $J_{n/2-k}%
\left(  z\right)  $ is convergent. Analytically continuing the integral
(\ref{int}) to $J_{\nu}\left(  z\right)  $, where $\nu$ can take on any
complex value, we achieve a finite result,%
\begin{align}
N\left(  \lambda;x,y\right)   &  =\sum_{k=0,\frac{1}{2},1,\cdots}^{\infty
}\frac{b_{k}\left(  x,y\right)  }{2^{k}\left(  2\pi\right)  ^{n/2}}\left(
\frac{\sqrt{\lambda-m^{2}}}{\left\vert x-y\right\vert }\right)  ^{n/2-k}%
J_{n/2-k}\left(  \sqrt{\lambda-m^{2}}\left\vert x-y\right\vert \right)
\nonumber\\
&  +\sum_{k=\frac{n}{2}+1,\frac{n}{2}+2,\cdots}^{\infty}\frac{b_{k}\left(
x,y\right)  }{\left(  4\pi\right)  ^{n/2}}\sum_{p=0}^{k-(n/2+1)}\frac{\left(
-1\right)  ^{p}}{2^{2p}p!}\left\vert x-y\right\vert ^{2p}\delta^{\left(
k-\left(  n/2+1\right)  -p\right)  }\left(  \lambda-m^{2}\right)
,\label{Nexpaned}%
\end{align}
where $\delta^{\left(  m\right)  }\left(  z\right)  =\frac{\partial^{m}%
}{\partial z^{m}}\delta\left(  z\right)  $. In this expansion, the terms with
$k<n/2+1$ are convergent and need not to be renormalized, and the terms with
$k\geq n/2+1$ are the renormalized terms.

The spectral counting function $N\left(  \lambda\right)  $ is the trace of
$N\left(  \lambda;x,y\right)  $:%
\begin{align}
N\left(  \lambda\right)   &  =\sum\limits_{k=0,\frac{1}{2},1,\cdots}^{\infty
}B_{k}\frac{\left(  \lambda-m^{2}\right)  ^{n/2-k}}{\left(  4\pi\right)
^{n/2}\Gamma\left(  n/2-k+1\right)  }+\sum\limits_{k=\frac{n}{2}+1,\frac{n}%
{2}+2,\cdots}^{\infty}B_{k}\frac{1}{\left(  4\pi\right)  ^{n/2}}%
\delta^{\left(  k-\left(  n/2+1\right)  \right)  }\left(  \lambda-m^{2}\right)
\nonumber\\
&  =\left(  \sum\limits_{k=0,\frac{1}{2},1,\cdots}^{n/2}+\sum\limits_{k=\frac
{n+1}{2},\frac{n+3}{2},\cdots}^{\infty}\right)  B_{k}\frac{\left(
\lambda-m^{2}\right)  ^{n/2-k}}{\left(  4\pi\right)  ^{n/2}\Gamma\left(
n/2-k+1\right)  }\nonumber\\
&  +\sum\limits_{k=\frac{n}{2}+1,\frac{n}{2}+2,\cdots}^{\infty}B_{k}\frac
{1}{\left(  4\pi\right)  ^{n/2}}\delta^{\left(  k-\left(  n/2+1\right)
\right)  }\left(  \lambda-m^{2}\right)  .
\end{align}
The case of $m=0$ recovers the result of Ref. \cite{Ours}, in which the
renormalization procedure is based on the analytical continuation of the gamma function.

\subsection{The series solution: general cases\label{expansion}}

To construct the general expansion for local counting functions, we perform
the transformation (\ref{KNq}) to the expansion of heat kernels, eq.
(\ref{GexpansionofKt}) and, then, we achieve%
\begin{align}
&  N\left(  \lambda;x,y\right) \nonumber\\
&  =\frac{1}{\left(  4\pi\right)  ^{n/2}}\left\{  \left[  \sum_{k=0,\frac
{1}{2},1,\cdots}^{N}b_{k}\left(  x,y\right)  +\sum_{k=N+\frac{1}{2}}^{\infty
}b_{k}^{\prime\prime}\left(  x,y\right)  +\sum_{k=N+\frac{1}{2}}^{\infty}%
b_{k}^{\prime}\left(  x,y\right)  \left[  \psi\left(  \frac{n}{2}+1-k\right)
-\ln\left(  \lambda-m^{2}\right)  \right]  \right]  \right. \nonumber\\
&  \times\left(  \frac{\left\vert x-y\right\vert }{2}\right)  ^{k-n/2}\left(
\lambda-m^{2}\right)  ^{n/4-k/2}J_{n/2-k}\left(  \sqrt{\lambda-m^{2}%
}\left\vert x-y\right\vert \right)  +\left[  \sum\limits_{k=\frac{n}%
{2}+1,\frac{n}{2}+2,\cdots}^{N}b_{k}\left(  x,y\right)  \right. \nonumber\\
&  +\left.  \sum\limits_{\substack{k=\frac{n}{2}+1,\frac{n}{2}+2,\cdots
\\k>N}}^{\infty}b_{k}^{\prime\prime}\left(  x,y\right)  \right]  \sum
_{p=0}^{k-\frac{n}{2}-1}\frac{\left(  -1\right)  ^{p}}{2^{2p}p!}\left\vert
x-y\right\vert ^{2p}\delta^{\left(  k-\left(  n/2+1\right)  -p\right)
}\left(  \lambda-m^{2}\right)  +\sum_{k=N+\frac{1}{2}}^{\infty}b_{k}^{\prime
}\left(  x,y\right) \nonumber\\
&  \times\left.  \Gamma\left(  \frac{n}{2}+1-k\right)  \left(  \lambda
-m^{2}\right)  ^{n/2-k}\left(  _{1}\tilde{F}_{2}\right)  _{a_{1}}\left(
\frac{n}{2}+1-k;\frac{n}{2}+1-k,\frac{n}{2}+1-k;-\frac{\left(  \lambda
-m^{2}\right)  \left(  x-y\right)  ^{2}}{4}\right)  \right\}  ,
\end{align}
where $\left(  _{1}\tilde{F}_{2}\right)  _{a_{1}}\left(  a_{1};b_{1}%
,b_{2};z\right)  =\frac{\partial}{\partial a_{1}}\left.  _{1}\tilde{F}%
_{2}\right.  \left(  a_{1};b_{1},b_{2};z\right)  $, $_{1}\tilde{F}_{2}\left(
a_{1};b_{1},b_{2};z\right)  =\frac{_{1}F_{2}\left(  a_{1};b_{1},b_{2}%
;z\right)  }{\Gamma\left(  b_{1}\right)  \Gamma\left(  b_{2}\right)  }$, and
$_{1}F_{2}\left(  a_{1};b_{1},b_{2};z\right)  $ is the generalized
hypergeometric function.

The expansion of the global counting function is the trace of $N\left(
\lambda;x,y\right)  $£º\begin{align}
N\left(  \lambda\right)   &  =\frac{1}{\left(  4\pi\right)  ^{n/2}}\left\{
\left(  \sum_{k=0,\frac{1}{2},1,\cdots}^{N}B_{k}+\sum_{k=N+\frac{1}{2}%
}^{\infty}B_{k}^{\prime\prime}\right)  \frac{\left(  \lambda-m^{2}\right)
^{n/2-k}}{\Gamma\left(  n/2+1-k\right)  }\right. \nonumber\\
&  +\left(  \sum\limits_{k=\frac{n}{2}+1,\frac{n}{2}+2,\cdots}^{N}B_{k}%
+\sum\limits_{\substack{k=\frac{n}{2}+1,\frac{n}{2}+2,\cdots\\k>N}}^{\infty
}B_{k}^{\prime\prime}\right)  \delta^{\left(  k-\left(  n/2+1\right)  \right)
}\left(  \lambda-m^{2}\right) \nonumber\\
&  \left.  +\sum_{k=N+\frac{1}{2}}^{\infty}B_{k}^{\prime}\frac{\left(
\lambda-m^{2}\right)  ^{n/2-k}}{\Gamma\left(  n/2+1-k\right)  }\left[
\psi\left(  \frac{n}{2}+1-k\right)  -\ln\left(  \lambda-m^{2}\right)  \right]
\right\}  .
\end{align}

\section{Scalar fields in $H_{3}$ (Euclidean $AdS_{3}$) and $H_{3}/Z$
(geometry of Euclidean BTZ black hole): one-loop effective actions, vacuum
energies, counting functions, and spectra\label{H3}}

In this section, we present the local and global regularized one-loop
effective actions, vacuum energies, counting functions, and spectra of scalar
fields in $H_{3}$ and $H_{3}/Z$. $H_{3}$, the three-dimensional hyperbolic
space, or, the Euclidean Anti-de Sitter space $AdS_{3}$, is a subspace of the
four-dimensional space with metric $ds^{2}=dX_{1}^{2}-dT_{1}^{2}+dX_{2}%
^{2}+dT_{2}^{2}$ satisfying the constraint $X_{1}^{2}-T_{1}^{2}+X_{2}%
^{2}+T_{2}^{2}=-l^{2}$ \cite{MSS}. $H_{3}/Z$ is the geometry of the Euclidean
BTZ black hole \cite{Carlip}, which is a quotient space of $H_{3}$ \cite{GMY}.
A clear description of $H_{3}$ and $H_{3}/Z$ can be found in Ref. \cite{GMY}.
Moreover, a series depth studies on spectral functions of hyperbolic spaces
are given in Refs. \cite{CH1,CH2,CH3,CH4}.

\subsection{The one-loop effective action in $H_{3}$}

For a scalar field in $H_{3}$, $D_{x}=-\nabla^{2}+m^{2}$ with $\nabla
^{2}=\partial_{r}^{2}+2\coth r\partial_{r}$, where $r\left(  x,y\right)
=\operatorname{arccosh}\left[  1+u\left(  x,y\right)  \right]  $ is the
geodesic distance between $x=\left(  \xi,\eta\right)  $ and $y=\left(
\xi^{\prime},\eta^{\prime}\right)  $ and $u\left(  x,y\right)  =\left[
\left(  \xi-\xi^{\prime}\right)  ^{2}+\left\vert \eta-\eta^{\prime}\right\vert
^{2}\right]  /\left(  2\xi\xi^{\prime}\right)  $ \cite{GMY}. The solution of
eq. (\ref{EqWdff}) gives the shifted local regularized one-loop effective
action:
\begin{equation}
W\left(  s;q;x,y\right)  =-\frac{\tilde{\mu}^{2s}}{2^{s+3/2}\pi^{3/2}}%
\frac{\left(  m^{2}+1+q\right)  ^{3/4-s/2}}{r^{1/2-s}\left(  x,y\right)  \sinh
r\left(  x,y\right)  }K_{3/2-s}\left(  \sqrt{m^{2}+1+q}r\left(  x,y\right)
\right)  .\label{LWH3}%
\end{equation}

The unshifted global regularized one-loop effective action can be achieved by
taking trace of $W\left(  s;0;x,y\right)  $. Dropping the divergent negative
power term gives the global regularized one-loop effective action,%
\begin{align}
W_{s}  &  =TrW\left(  s;0;x,y\right) \nonumber\\
&  =-Vol\left(  H_{3}\right)  \frac{\tilde{\mu}^{2s}}{16\pi^{3/2}}%
\Gamma\left(  s-\frac{3}{2}\right)  \left(  m^{2}+1\right)  ^{3/2-s}.
\end{align}
Here $s=0$ is not a singular point and the regularized one-loop effective
action without the regularization parameter $s$ is just $\left.
W_{s}\right\vert _{s=0}$:
\begin{equation}
W=-Vol\left(  H_{3}\right)  \frac{1}{12\pi}\left(  m^{2}+1\right)  ^{3/2}.
\end{equation}
This agrees with the result given by Ref. \cite{GMY}.

\subsection{The vacuum energy in $H_{3}$}

With $D_{x}=-\nabla^{2}+m^{2}$ and $\nabla^{2}=\partial_{r}^{2}+2\coth
r\partial_{r}$ \cite{GMY}, the solution of eq. (\ref{EqofveDeq}) gives the
shifted local regularized vacuum energy in $H_{3}$,%
\begin{equation}
E_{0}\left(  \epsilon;q;x,y\right)  =\frac{\tilde{\mu}^{2\epsilon}%
}{2^{1+\epsilon}\pi^{3/2}}\frac{1}{\Gamma\left(  -1/2+\epsilon\right)  }%
\frac{\left(  m^{2}+1+q\right)  ^{1-\epsilon/2}}{r\left(  x,y\right)
^{1-\epsilon}\sinh r\left(  x,y\right)  }K_{2-\epsilon}\left(  \sqrt
{m^{2}+1+q}r\left(  x,y\right)  \right)  .
\end{equation}
The global regularized vacuum energy is the trace of $E_{0}\left(
\epsilon;0;x,y\right)  $:%
\begin{align}
E_{0}\left(  \epsilon;0\right)   &  =TrE_{0}\left(  \epsilon;0;x,y\right)
\nonumber\\
&  =Vol\left(  H_{3}\right)  \frac{\tilde{\mu}^{2\epsilon}}{16\pi^{3/2}}%
\frac{\Gamma\left(  -2+\epsilon\right)  }{\Gamma\left(  -1/2+\epsilon\right)
}\left(  m^{2}+1\right)  ^{2-\epsilon}.
\end{align}
Laurent expanding $E_{0}\left(  \epsilon;0\right)  $,%
\begin{align}
&  E_{0}\left(  \epsilon\right)  =-Vol\frac{\left(  m^{2}+1\right)  ^{2}%
}{64\pi^{2}}\nonumber\\
&  \times\left\{  \frac{1}{\epsilon}+\left(  \ln\frac{4\tilde{\mu}^{2}}%
{m^{2}+1}-\frac{1}{2}\right)  +\epsilon\left[  \frac{1}{2}\left(  \ln
\frac{4\tilde{\mu}^{2}}{m^{2}+1}\right)  ^{2}-\frac{1}{2}\ln\frac{4\tilde{\mu
}^{2}}{m^{2}+1}-\frac{5}{4}-\frac{\pi^{2}}{6}\right]  \right. \nonumber\\
&  \left.  +\epsilon^{2}\left[  \frac{1}{6}\left(  \ln\frac{4\tilde{\mu}^{2}%
}{m^{2}+1}\right)  ^{3}-\frac{1}{4}\left(  \ln\frac{4\tilde{\mu}^{2}}{m^{2}%
+1}\right)  ^{2}-\left(  \frac{5}{4}+\frac{\pi^{2}}{6}\right)  \ln
\frac{4\tilde{\mu}^{2}}{m^{2}+1}-\frac{13}{8}+\frac{\pi^{2}}{12}+2\zeta\left(
3\right)  \right]  +\cdots\right\}  ,
\end{align}
taking $\epsilon=0$, and dropping the divergent negative power term, we
achieve a regularized vacuum energy without the regularization parameter
$\epsilon$,%
\begin{equation}
E_{0}=Vol\left(  H_{3}\right)  \frac{\left(  m^{2}+1\right)  ^{2}}{64\pi^{2}%
}\left(  \frac{1}{2}+\ln\frac{m^{2}+1}{4\tilde{\mu}^{2}}\right)  .
\end{equation}

\subsection{The counting function and the spectrum in $H_{3}$}

The solution of eq. (\ref{dieqofN}) gives the local counting function in
$H_{3}$,%
\begin{equation}
N\left(  \lambda;x,y\right)  =\frac{\sin\left(  \sqrt{\lambda-\left(
m^{2}+1\right)  }r\left(  x,y\right)  \right)  -r\left(  x,y\right)
\sqrt{\lambda-\left(  m^{2}+1\right)  }\cos\left(  \sqrt{\lambda-\left(
m^{2}+1\right)  }r\left(  x,y\right)  \right)  }{2\pi^{2}r\left(  x,y\right)
^{2}\sinh r\left(  x,y\right)  }.\label{NH3}%
\end{equation}
Taking trace gives the global counting function,%
\begin{equation}
N\left(  \lambda\right)  =Vol\left(  H_{3}\right)  \frac{\left[
\lambda-\left(  m^{2}+1\right)  \right]  ^{3/2}}{6\pi^{2}}.
\end{equation}

From a counting function, one can immediately achieve the eigenvalue spectrum
of the operator $D$ \cite{Ours}. By $N\left(  \lambda_{n}\right)  =n$, we can
obtain the spectrum,%
\begin{equation}
\lambda_{n}=m^{2}+1+\left[  \frac{6\pi^{2}n}{Vol\left(  H_{3}\right)
}\right]  ^{2/3}.
\end{equation}

\subsection{The one-loop effective action in $H_{3}/Z$}

$H_{3}/Z$ is a quotient of $H_{3}$. The solution in the quotient space
$H_{3}/Z$ can be represented as a linear combination of the solutions in the
space $H_{3}$ once the equation is linear \cite{GMY,Schulman}. Concretely, for
one-loop effective actions, the solution of eq. (\ref{EqW}) on $V/\Gamma$ can
be expressed as a linear combination of the solutions of eq. (\ref{EqWdff}) on
$V$:%
\begin{equation}
W^{V/\Gamma}\left(  s;q;x,y\right)  =\sum_{\alpha\in\Gamma}W^{V}\left(
s;q;x,\alpha y\right)  .
\end{equation}
Then for $H_{3}$ and its quotient $H_{3}/Z$, we have%
\begin{equation}
W^{H_{3}/Z}\left(  s;q;x,y\right)  =\sum_{n=-\infty}^{\infty}W^{H_{3}}\left(
s;q;x,\gamma^{n}y\right)  ,
\end{equation}
where $y=\left(  \xi,\eta\right)  $ and $\gamma y=\gamma\left(  \xi
,\eta\right)  \rightarrow\left(  \left\vert \lambda\right\vert ^{-1}%
\xi,\lambda^{-1}\eta\right)  $ with $\lambda=e^{i2\pi\tau}$ and $\tau=\left(
\theta+i\beta\right)  /2\pi$; here $\beta$ is the temperature and $\theta$ is
the angular potential of a thermal Anti-de Sitter space \cite{GMY}.

In the present case, the shifted local regularized one-loop effective action
of a scalar field in space $H_{3}/Z$ is%
\begin{equation}
W^{H_{3}/Z}\left(  s;q;x,y\right)  =-\frac{\tilde{\mu}^{2s}}{2^{3/2+s}%
\pi^{3/2}}\sum_{n=-\infty}^{\infty}\left(  \sqrt{m^{2}+1+q}\right)
^{3/2-s}\frac{K_{3/2-s}\left(  \sqrt{m^{2}+1+q}r\left(  x,\gamma^{n}y\right)
\right)  }{r\left(  x,\gamma^{n}y\right)  ^{1/2-s}\sinh r\left(  x,\gamma
^{n}y\right)  }.
\end{equation}

The corresponding global regularized one-loop effective action can be achieved
by taking trace of $W^{H_{3}/Z}\left(  s;0;x,y\right)  $:%
\begin{align}
W_{s}  &  =TrW\left(  s;0;x,y\right)  =\int_{H_{3}/Z}d^{3}x\sqrt{g}W\left(
s;0;x,x\right) \nonumber\\
&  =-Vol\left(  H_{3}/Z\right)  \frac{\tilde{\mu}^{2s}}{16\pi^{3/2}}%
\Gamma\left(  s-\frac{3}{2}\right)  \left(  \sqrt{m^{2}+1}\right)
^{3-s/2}\nonumber\\
&  -\frac{\tilde{\mu}^{2s}\beta^{1/2+s}}{2^{1/2+s}\sqrt{\pi}}\left(
\sqrt{m^{2}+1}\right)  ^{1/2-s}\sum_{n=1}^{\infty}\frac{K_{1/2-s}\left(
\sqrt{m^{2}+1}n\beta\right)  }{n^{1/2-s}\left[  \cosh\left(  n\beta\right)
-\cos\left(  n\theta\right)  \right]  },\label{WsH3Z}%
\end{align}
where we have used $\int_{H_{3}/Z}d^{3}x\sqrt{g}=\int_{n\beta}^{\infty}%
dr\int_{1}^{e^{\beta}}d\rho\int_{0}^{2\pi}d\phi\frac{1}{\rho}\frac{\sinh
r}{2\left[  \cosh\left(  n\beta\right)  -\cos\left(  n\theta\right)  \right]
}$. This agrees with the result given by Ref. \cite{GMY}.

Here $s=0$ is not a singular point, so the one-loop effective action can be
directly obtained by substituting $s=0$ into eq. (\ref{WsH3Z}):
\begin{equation}
W=-Vol\left(  H_{3}/Z\right)  \frac{1}{12\pi}\left(  m^{2}+1\right)
^{3/2}-\sqrt{\frac{\beta}{2\pi}}\left(  m^{2}+1\right)  ^{1/4}\sum
_{n=1}^{\infty}\frac{K_{1/2}\left(  \sqrt{m^{2}+1}n\beta\right)  }{\sqrt
{n}\left[  \cosh\left(  n\beta\right)  -\cos\left(  n\theta\right)  \right]
}.
\end{equation}

\subsection{The vacuum energy in $H_{3}/Z$}

Based on the result of the vacuum energy in $H_{3}$, using the treatment that
we have used in obtaining the one-loop effective action in $H_{3}/Z$, we can
achieve the local vacuum energy in $H_{3}/Z$,%
\begin{equation}
E_{0}\left(  \epsilon;q;x,y\right)  =\frac{\tilde{\mu}^{2\epsilon}%
}{2^{\epsilon}\pi^{3/2}\Gamma\left(  -1/2+\epsilon\right)  }\left(
m^{2}+1+q\right)  ^{1-\epsilon/2}\sum_{n=-\infty}^{\infty}\frac{K_{2-\epsilon
}\left(  \sqrt{m^{2}+1+q}r\left(  x,\gamma^{n}y\right)  \right)  }{r\left(
x,\gamma^{n}y\right)  ^{1-\epsilon}\sinh r\left(  x,\gamma^{n}y\right)  }.
\end{equation}
Taking trace of $E_{0}\left(  \epsilon;0;x,y\right)  $ gives the global vacuum
energy,
\begin{align}
E_{0}\left(  \epsilon\right)   &  =TrE_{0}\left(  \epsilon;0;x,y\right)
=\int_{H_{3}/Z}d^{3}x\sqrt{g}E_{0}\left(  \epsilon;0;x,x\right) \nonumber\\
&  =Vol\left(  H_{3}/Z\right)  \frac{\tilde{\mu}^{2\epsilon}}{16\pi^{3/2}%
}\frac{\Gamma\left(  -2+\epsilon\right)  }{\Gamma\left(  -1/2+\epsilon\right)
}\left(  m^{2}+1\right)  ^{2-\epsilon}\nonumber\\
&  +\sum_{n=1}^{\infty}\frac{\tilde{\mu}^{2\epsilon}\beta^{\epsilon}%
}{2^{\epsilon}\pi^{1/2}\Gamma\left(  -1/2+\epsilon\right)  }\left(
\sqrt{m^{2}+1}\right)  ^{1-\epsilon}\frac{K_{1-\epsilon}\left(  n\beta
\sqrt{m^{2}+1}\right)  }{n^{1-\epsilon}\left[  \cosh\left(  n\beta\right)
-\cos\left(  n\theta\right)  \right]  }.
\end{align}
Laurent expanding $E_{0}\left(  \epsilon\right)  $,%
\begin{align}
&  E_{0}\left(  \epsilon\right) \nonumber\\
&  =-\frac{1}{\epsilon}Vol\frac{\left(  m^{2}+1\right)  ^{2}}{64\pi^{2}%
}\nonumber\\
&  -\left\{  Vol\frac{\left(  m^{2}+1\right)  ^{2}}{64\pi^{2}}\left(  \ln
\frac{4\tilde{\mu}^{2}}{m^{2}+1}-\frac{1}{2}\right)  +\sum_{n=1}^{\infty}%
\frac{\sqrt{m^{2}+1}}{4n\pi\left\vert \sin\left(  n\pi\tau\right)  \right\vert
^{2}}K_{1}\left(  2n\pi\tau_{2}\sqrt{m^{2}+1}\right)  \right\} \nonumber\\
&  -\epsilon\left\{  Vol\frac{\left(  m^{2}+1\right)  ^{2}}{64\pi^{2}}\left[
\frac{1}{2}\left(  \ln\frac{4\tilde{\mu}^{2}}{m^{2}+1}\right)  ^{2}-\frac
{1}{2}\ln\frac{4\tilde{\mu}^{2}}{m^{2}+1}-\frac{5}{4}-\frac{\pi^{2}}%
{6}\right]  \right. \nonumber\\
&  \left.  +\sum_{n=1}^{\infty}\frac{\sqrt{m^{2}+1}}{4n\pi\left\vert
\sin\left(  n\pi\tau\right)  \right\vert ^{2}}\left[  \left(  \ln\frac
{4n\pi\tau_{2}\tilde{\mu}^{2}}{\sqrt{m^{2}+1}}+\gamma_{E}-2\right)
K_{1}\left(  2n\pi\tau_{2}\sqrt{m^{2}+1}\right)  -K_{1}^{\left(  1\right)
}\left(  2n\pi\tau_{2}\sqrt{m^{2}+1}\right)  \right]  \right\} \nonumber\\
&  -\epsilon^{2}\left\{  Vol\frac{\left(  m^{2}+1\right)  ^{2}}{64\pi^{2}%
}\left[  \frac{1}{6}\left(  \ln\frac{4\tilde{\mu}^{2}}{m^{2}+1}\right)
^{3}-\frac{1}{4}\left(  \ln\frac{4\tilde{\mu}^{2}}{m^{2}+1}\right)
^{2}-\left(  \frac{5}{4}+\frac{\pi^{2}}{6}\right)  \ln\frac{4\tilde{\mu}^{2}%
}{m^{2}+1}-\frac{13}{8}+\frac{\pi^{2}}{12}+2\zeta\left(  3\right)  \right]
\right. \nonumber\\
&  +\frac{1}{4}\sum_{n=1}^{\infty}\frac{\sqrt{m^{2}+1}}{4n\pi\left\vert
\sin\left(  n\pi\tau\right)  \right\vert ^{2}}\left\{  2K_{1}^{\left(
2\right)  }\left(  2n\pi\tau_{2}\sqrt{m^{2}+1}\right)  -\frac{2\left(
\ln\frac{4n\pi\tau_{2}\tilde{\mu}^{2}}{\sqrt{m^{2}+1}}+\gamma_{E}-2\right)
}{n\pi\tau_{2}\sqrt{m^{2}+1}}K_{0}\left(  2n\pi\tau_{2}\sqrt{m^{2}+1}\right)
\right. \nonumber\\
&  \left.  \left.  +\left[  2\ln\frac{4n\pi\tau_{2}\tilde{\mu}^{2}}%
{\sqrt{m^{2}+1}}\left(  \ln\frac{4n\pi\tau_{2}\tilde{\mu}^{2}}{\sqrt{m^{2}+1}%
}+2\gamma_{E}-4\right)  +2\gamma_{E}\left(  \gamma_{E}-4\right)  -\pi
^{2}\right]  K_{1}\left(  2n\pi\tau_{2}\sqrt{m^{2}+1}\right)  \right\}
\right\}  +\cdots,
\end{align}
taking $\epsilon=0$, and dropping the divergent negative power term gives a
regularized vacuum energy without the regularization parameter $\epsilon$,%
\begin{equation}
E_{0}=Vol\left(  H_{3}/Z\right)  \frac{\left(  m^{2}+1\right)  ^{2}}{64\pi
^{2}}\left(  \frac{1}{2}+\ln\frac{m^{2}+1}{4\tilde{\mu}^{2}}\right)
-\sum_{n=1}^{\infty}\sqrt{m^{2}+1}\frac{K_{1}\left(  n\beta\sqrt{m^{2}%
+1}\right)  }{2\pi n\left[  \cosh\left(  n\beta\right)  -\cos\left(
n\theta\right)  \right]  }.
\end{equation}

\subsection{The counting function and the spectrum in $H_{3}/Z$}

From the local counting function in $H_{3}$, eq. (\ref{NH3}), we can obtain
the local counting function in $H_{3}/Z$,%
\begin{align}
&  N\left(  \lambda;x,y\right) \nonumber\\
&  =\sum\limits_{n=-\infty}^{\infty}\frac{\sin\left(  \sqrt{\lambda-\left(
m^{2}+1\right)  }r\left(  x,\gamma^{n}y\right)  \right)  -\sqrt{\lambda
-\left(  m^{2}+1\right)  }r\left(  x,\gamma^{n}y\right)  \cos\left(
\sqrt{\lambda-\left(  m^{2}+1\right)  }r\left(  x,\gamma^{n}y\right)  \right)
}{2\pi^{2}r^{2}\left(  x,\gamma^{n}y\right)  \sinh r\left(  x,\gamma
^{n}y\right)  }.
\end{align}

Taking trace gives the global counting function,%
\begin{equation}
N\left(  \lambda\right)  =Vol\left(  H_{3}/Z\right)  \frac{\left[
\lambda-\left(  m^{2}+1\right)  \right]  ^{3/2}}{6\pi^{2}}+\sum_{n=1}^{\infty
}\frac{\sin\left(  \sqrt{\lambda-\left(  m^{2}+1\right)  }n\beta\right)  }{\pi
n\left[  \cosh\left(  n\beta\right)  -\cos\left(  n\theta\right)  \right]
}.\label{H3ZNLmd}%
\end{equation}

The sum in eq. (\ref{H3ZNLmd}) is difficult to be solved, so we turn to
consider an approximate result. When $\beta$ is very large, only the first
term, corresponding to $n=1$ is important. Then we approximately achieve
\begin{equation}
N\left(  \lambda\right)  \approx Vol\left(  H_{3}/Z\right)  \frac{\left[
\lambda-\left(  m^{2}+1\right)  \right]  ^{3/2}}{6\pi^{2}}+\frac{1}{\pi}%
\frac{\sin\left(  \sqrt{\lambda-\left(  m^{2}+1\right)  }\beta\right)  }%
{\cosh\beta-\cos\theta}.
\end{equation}
The spectrum is determined by $N\left(  \lambda_{n}\right)  =n$; approximately
solving this equation gives
\begin{equation}
\lambda_{n}\approx m^{2}+1+\left[  \frac{6\pi^{2}n}{Vol\left(  H_{3}/Z\right)
}\right]  ^{2/3}\left(  1-\frac{2\sin\alpha}{3n\pi\cosh\beta-3n\pi\cos
\theta+\alpha\cos\alpha}\right)  ,
\end{equation}
where $\alpha=\left[  6\pi^{2}n/Vol\left(  H_{3}/Z\right)  \right]
^{1/3}\beta$.

\section{Massless scalar fields in $S^{1}$: one-loop effective actions, vacuum
energies, counting functions, and spectra \label{S1}}

In this section, we consider the local and global one-loop effective actions,
vacuum energies, counting functions, and spectra of a massless scalar field in
$S^{1}=\mathbb{R}^{1}/Z$, the quotient of $\mathbb{R}^{1}$.

\subsection{The one-loop effective action}

As discussed above, the shifted local regularized one-loop effective action,
the solution of eq. (\ref{EqWdff}), in the quotient space $\mathbb{R}^{1}/Z$
is the linear combination of the solutions in space $\mathbb{R}^{1}$. The
solution in space $\mathbb{R}^{1}$ can be solved from eq. (\ref{EqWdff}):
$W^{\mathbb{R}^{1}}\left(  s;q;x,y\right)  =-\frac{1}{2^{s}\sqrt{2\pi}}%
\tilde{\mu}^{2s}\left(  \sqrt{q}/\left\vert x-y\right\vert \right)
^{1/2-s}K_{1/2-s}\left(  \sqrt{q}\left\vert x-y\right\vert \right)  $. Then
the shifted local regularized one-loop effective action reads:%
\begin{align}
W\left(  s;q;x,y\right)   &  =\sum_{Z}W^{\mathbb{R}^{1}}\left(  s;q;x,y\right)
\nonumber\\
&  =-\frac{q^{1/2-s}\tilde{\mu}^{2s}}{2^{s+1/2}\sqrt{\pi}}\sum_{n=-\infty
}^{\infty}\frac{K_{1/2-s}\left(  \sqrt{q}\left\vert y+nL-x\right\vert \right)
}{\left(  \sqrt{q}\left\vert y+nL-x\right\vert \right)  ^{1/2-s}},
\end{align}
where $L$ is the perimeter of $S^{1}$.

The shifted global regularized one-loop effective action can be directly
achieved by taking trace. After dropping the divergent power term, we have%
\begin{align}
W\left(  s;q\right)   &  =TrW\left(  s;q;x,y\right) \nonumber\\
&  =-L\frac{q^{1/2-s}\tilde{\mu}^{2s}}{4\sqrt{\pi}}\left[  \Gamma\left(
s-\frac{1}{2}\right)  +2^{5/2-s}\sum_{n=1}^{\infty}\frac{K_{1/2-s}\left(
n\sqrt{q}L\right)  }{\left(  n\sqrt{q}L\right)  ^{1/2-s}}\right]  .
\end{align}
The unshifted regularized one-loop effective action is
\begin{equation}
W_{s}=\lim_{q\rightarrow0}W\left(  s;q\right)  =-\left(  \frac{\tilde{\mu}%
L}{2}\right)  ^{2s}\frac{1}{\sqrt{\pi}}\Gamma\left(  \frac{1}{2}-s\right)
\zeta\left(  1-2s\right)  .
\end{equation}
The regularized one-loop effective action can be obtained by Laurent expanding
$W_{s}$ around $s=0$,%
\begin{align}
W_{s}  &  =\frac{1}{2s}-\frac{1}{2}\gamma_{E}+\ln L\tilde{\mu}\nonumber\\
&  +\frac{1}{2\sqrt{\pi}}\sum_{p=2}^{\infty}\left[  \sum_{\beta=0}^{p}%
\frac{\left(  -1\right)  ^{p-\beta}\Gamma^{\left(  p-\beta\right)  }\left(
1/2\right)  }{\beta!\left(  p-\beta\right)  !}\left(  \ln\frac{L^{2}\tilde
{\mu}^{2}}{4}\right)  ^{\beta}\right]  s^{p-1}\nonumber\\
&  -\frac{1}{\sqrt{\pi}}\sum_{p=1}^{\infty}\left[  \sum_{\beta=0}^{p}%
\sum_{\rho=0}^{p-\beta}\frac{\left(  -1\right)  ^{\rho}2^{p-\beta-\rho}%
\Gamma^{\left(  \rho\right)  }\left(  1/2\right)  \gamma_{p-\beta-\rho}}%
{\beta!\rho!\left(  p-\beta-\rho\right)  !}\left(  \ln\frac{L^{2}\tilde{\mu
}^{2}}{4}\right)  ^{\beta}\right]  s^{p}.
\end{align}
Taking $s=0$ and dropping the divergent negative power term gives the
regularized result without the regularization parameter $\epsilon$,%
\begin{equation}
W=-\frac{1}{2}\gamma_{E}+\ln\left(  \tilde{\mu}L\right)  .
\end{equation}

Moreover, if we approximately replace the sum $\sum_{n=1}^{\infty}$by the
integral $%
%TCIMACRO{\dint _{1}^{\infty}}%
%BeginExpansion
{\displaystyle\int_{1}^{\infty}}
%EndExpansion
dn$, the shifted local and global regularized one-loop effective actions can
be calculated analytically, respectively,%
\begin{align}
&  W\left(  s;q;x,y\right) \nonumber\\
&  \simeq-\frac{1}{2L}\frac{\tilde{\mu}^{2s}}{q^{s}}\Gamma\left(  s\right)
+\frac{1}{4\sqrt{\pi}L}\frac{\tilde{\mu}^{2s}}{q^{s}}\left\{  2\Gamma\left(
s-\frac{1}{2}\right)  \left[  \sqrt{q}\frac{y-x}{2}\left.  _{1}F_{2}\right.
\left(  \frac{1}{2};\frac{3}{2},\frac{3}{2}-s;q\frac{\left(  y-x\right)  ^{2}%
}{4}\right)  \right.  \right. \nonumber\\
&  \left.  +\sqrt{q}\frac{L-\left(  y-x\right)  }{2}\left.  _{1}F_{2}\right.
\left(  \frac{1}{2};\frac{3}{2},\frac{3}{2}-s;q\frac{\left[  L-\left(
y-x\right)  \right]  ^{2}}{4}\right)  \right]  +\frac{1}{s}\Gamma\left(
\frac{1}{2}-s\right)  \left\{  \left(  \sqrt{q}\frac{y-x}{2}\right)
^{2s}\right. \nonumber\\
&  \times\left.  _{1}F_{2}\right.  \left(  s;s+\frac{1}{2},s+1;\frac{q\left(
y-x\right)  ^{2}}{4}\right)  \left.  \left.  +\left[  \sqrt{q}\frac{L-\left(
y-x\right)  }{2}\right]  ^{2s}\left.  _{1}F_{2}\right.  \left(  s;s+\frac
{1}{2},s+1;q\frac{\left[  L-\left(  y-x\right)  \right]  ^{2}}{4}\right)
\right\}  \right\}
\end{align}
and%
\begin{align}
W\left(  s;q\right)   &  \simeq-L\frac{\tilde{\mu}^{2s}q^{1/2-s}}{4\sqrt{\pi}%
}\Gamma\left(  s-\frac{1}{2}\right)  -\frac{\tilde{\mu}^{2s}}{2q^{s}}%
\Gamma\left(  s\right)  +\frac{\tilde{\mu}^{2s}}{\sqrt{\pi}q^{s}}\left[
\frac{\sqrt{q}L}{2}\Gamma\left(  s-\frac{1}{2}\right)  \left.  _{1}%
F_{2}\right.  \left(  \frac{1}{2};\frac{3}{2},\frac{3}{2}-s;\left(
\frac{\sqrt{q}L}{2}\right)  ^{2}\right)  \right. \nonumber\\
&  +\left.  \left(  \frac{\sqrt{q}L}{2}\right)  ^{2s}\frac{1}{2s}\Gamma\left(
\frac{1}{2}-s\right)  \left.  _{1}F_{2}\right.  \left(  s;s+\frac{1}%
{2},s+1;\left(  \frac{\sqrt{q}L}{2}\right)  ^{2}\right)  \right]  ,
\end{align}
where we assume that $y\geq x$,

\subsection{The vacuum energy}

The shifted local regularized vacuum energy of a massless scalar field in
$\mathbb{R}^{1}$ can be solved from eq. (\ref{EqofveDeq}), $E_{0}%
^{\mathbb{R}^{1}}\left(  \epsilon;q;x,y\right)  =\frac{\tilde{\mu}^{2\epsilon
}}{\Gamma\left(  -1/2+\epsilon\right)  }\frac{q^{1-\epsilon}}{2^{\epsilon
}\sqrt{\pi}}K_{1-\epsilon}\left(  \sqrt{q}\left\vert y-x\right\vert \right)
/\left(  \sqrt{q}\left\vert y-x\right\vert \right)  ^{1-\epsilon}$. Then, the
shifted local regularized vacuum energy in $S^{1}$ is%
\begin{equation}
E_{0}\left(  \epsilon;q;x,y\right)  =\frac{\tilde{\mu}^{2\epsilon
}q^{1-\epsilon}}{2^{\epsilon}\sqrt{\pi}\Gamma\left(  -1/2+\epsilon\right)
}\sum_{n=-\infty}^{\infty}\frac{K_{1-\epsilon}\left(  \sqrt{q}\left\vert
y+nL-x\right\vert \right)  }{\left(  \sqrt{q}\left\vert y+nL-x\right\vert
\right)  ^{1-\epsilon}},\label{veS1local}%
\end{equation}
since $S^{1}$ is the quotient space of $\mathbb{R}^{1}$.

The shifted global regularized vacuum energy can be obtained by taking trace:%
\begin{align}
E_{0}\left(  \epsilon;q\right)   &  =TrE_{0}\left(  \epsilon;q;x,y\right)
\nonumber\\
&  =L\frac{\tilde{\mu}^{2\epsilon}}{\Gamma\left(  -1/2+\epsilon\right)  }%
\frac{q^{1-\epsilon}}{2^{\epsilon}\sqrt{\pi}}\left[  \frac{\Gamma\left(
-1+\epsilon\right)  }{2^{2-\epsilon}}+2\sum_{n=1}^{\infty}\frac{K_{1-\epsilon
}\left(  \sqrt{q}nL\right)  }{\left(  \sqrt{q}nL\right)  ^{1-\epsilon}%
}\right]  .
\end{align}
Laurent expanding $E_{0}\left(  \epsilon;q\right)  $ with respect to
$\epsilon$ gives%
\begin{align}
E_{0}\left(  \epsilon;q\right)   &  =\frac{1}{\epsilon}\frac{qL}{8\pi}%
+\frac{qL}{8\pi}\left(  \ln\frac{4\tilde{\mu}^{2}}{q}-1\right)  -\sum
_{n=1}^{\infty}\frac{\sqrt{q}K_{1}\left(  \sqrt{q}nL\right)  }{n\pi
}\nonumber\\
&  +\epsilon\left\{  \frac{qL}{48\pi}\left[  3\left(  \ln\frac{4\tilde{\mu
}^{2}}{q}\right)  ^{2}-6\ln\frac{4\tilde{\mu}^{2}}{q}-6-\pi^{2}\right]
\right. \nonumber\\
&  \left.  +\sum_{n=1}^{\infty}\frac{1}{\pi Ln^{2}}\left[  K_{0}\left(
\sqrt{q}nL\right)  -\sqrt{q}nLK_{1}\left(  \sqrt{q}nL\right)  \left(  \ln
\frac{2nL\tilde{\mu}^{2}}{\sqrt{q}}+\gamma_{E}-2\right)  \right]  \right\}
+\cdots.
\end{align}
Taking $\epsilon=0$ and dropping the divergent negative power term gives the
regularized shifted global vacuum energy without the regularization parameter
$\epsilon$,%
\begin{equation}
E_{0}\left(  0;q\right)  =-\frac{qL}{8\pi}\left(  1+\ln\frac{q}{4\tilde{\mu
}^{2}}\right)  -\frac{qL}{\pi}\sum_{n=1}^{\infty}\frac{K_{1}\left(  \sqrt
{q}nL\right)  }{\sqrt{q}nL}.
\end{equation}
The unshifted vacuum energy is then given by taking the limit $q\rightarrow0$,%
\begin{equation}
E_{0}=-\sum_{n=1}^{\infty}\frac{1}{\pi Ln^{2}}=-\frac{\pi}{6L}.
\end{equation}

By the way, by approximately replacing the sum $\sum_{n=1}^{\infty}$ with the
integral $%
%TCIMACRO{\dint _{1}^{\infty}}%
%BeginExpansion
{\displaystyle\int_{1}^{\infty}}
%EndExpansion
dn$, we can achieve an analytical expression of the shifted local regularized
vacuum energy,%
\begin{align}
&  E\left(  \epsilon;q;x,y\right) \nonumber\\
&  \simeq\frac{\tilde{\mu}^{2\epsilon}q^{1/2-\epsilon}}{2L}+\frac{\tilde{\mu
}^{2\epsilon}}{\Gamma\left(  -1/2+\epsilon\right)  }\frac{q^{1/2-\epsilon}%
}{2^{\epsilon}\sqrt{\pi}L}\left\{  \frac{\Gamma\left(  1-\epsilon\right)
}{\left(  1-2\epsilon\right)  2^{1-\epsilon}}\right. \nonumber\\
&  \times\left\{  \frac{_{1}F_{2}\left(  \epsilon-1/2;\epsilon,\epsilon
+1/2;q\left(  y-x\right)  ^{2}/4\right)  }{\left[  \sqrt{q}\left(  y-x\right)
/2\right]  ^{1-2\epsilon}}+\frac{_{1}F_{2}\left(  \epsilon-1/2;\epsilon
,\epsilon+1/2;q\left[  L-\left(  y-x\right)  \right]  ^{2}/4\right)  }{\left[
\sqrt{q}\left(  L-\left(  y-x\right)  \right)  /2\right]  ^{1-2\epsilon}%
}\right\} \nonumber\\
&  -\frac{\Gamma\left(  \epsilon-1\right)  }{2^{1-\epsilon}}\left\{
\frac{\sqrt{q}\left(  y-x\right)  }{2}\left.  _{1}F_{2}\right.  \left(
\frac{1}{2};\frac{3}{2},2-\epsilon;\frac{q\left(  y-x\right)  ^{2}}{4}\right)
\right. \nonumber\\
&  \left.  \left.  +\frac{\sqrt{q}\left[  L-\left(  y-x\right)  \right]  }%
{2}\left.  _{1}F_{2}\right.  \left(  \frac{1}{2};\frac{3}{2},2-\epsilon
;\frac{q\left[  L-\left(  y-x\right)  \right]  ^{2}}{4}\right)  \right\}
\right\}  .
\end{align}
The corresponding regularized vacuum energy is%
\begin{equation}
E_{0}\left(  0;q\right)  \simeq-\frac{qL}{8\pi}\left[  \left(  1+\ln\frac
{q}{4\tilde{\mu}^{2}}\right)  \right]  -\frac{\sqrt{q}}{4\pi}G_{1,3}%
^{3,0}\left(  \left.  \frac{qL^{2}}{4}\right\vert
\begin{array}
[c]{c}%
1\\
\frac{1}{2},0,-\frac{1}{2}%
\end{array}
\right)  ,
\end{equation}
where $G_{p,q}^{m,n}\left(  z\left\vert
\begin{array}
[c]{c}%
a_{1},\cdots a_{n},a_{n+1},\cdots,a_{p}\\
b_{1},\cdots b_{m},b_{m+1},\cdots,b_{q}%
\end{array}
\right.  \right)  $ is Meijer's G-function. Then, $q\rightarrow0$ gives%
\begin{equation}
E_{0}\simeq-\frac{1}{\pi L}.
\end{equation}

\subsection{The counting function and the spectrum}

The local counting function of a massless scalar field in $\mathbb{R}^{1}$ can
be solved from eq. (\ref{eqofNdiff}), $N\left(  \lambda;x,y\right)  =\frac
{1}{\pi}\sin\left(  \sqrt{\lambda}\left\vert y-x\right\vert \right)
/\left\vert y-x\right\vert $. Then, the local counting function in $S^{1}$, a
quotient space of $\mathbb{R}^{1}$, reads%
\begin{equation}
N\left(  \lambda;x,y\right)  =\frac{1}{\pi}\sum_{n=-\infty}^{\infty}\frac
{\sin\left(  \sqrt{\lambda}\left\vert y+nL-x\right\vert \right)  }{\left\vert
y+nL-x\right\vert }.
\end{equation}
The global counting function $N\left(  \lambda\right)  $ is the trace of
$N\left(  \lambda;x,y\right)  $,
\begin{equation}
N\left(  \lambda\right)  =trN\left(  \lambda;x,y\right)  =2k+1,\text{
\ }\left(  \frac{2k\pi}{L}\right)  ^{2}<\lambda<\left[  \frac{2\left(
k+1\right)  \pi}{L}\right]  ^{2},\text{ \ }k=0,1,2,\cdots
.\label{LocalNLambdaS1}%
\end{equation}

From the counting function, we can achieve the eigenvalue spectrum of the
operator $D$:
\begin{align}
\lambda_{0}  &  =0,\nonumber\\
\lambda_{2k+1}  &  =\lambda_{2k+2}=\left[  \frac{2\left(  k+1\right)  \pi}%
{L}\right]  ^{2},\text{ \ }k=0,1,2,\cdots.
\end{align}

\section{The Higgs model in a $(1+1)$-dimensional finite interval with the
Dirichlet boundary condition: one-loop effective actions, vacuum energies,
counting functions, and spectra\label{Higgs}}

In the $(1+1)$-dimensional Higgs model \cite{GMS}, we concern ourselves with
the fluctuation $\delta H\left(  x^{\mu}\right)  $ which is a linearized shift
of a scalar field from the homogeneous stable solution. Here, the shift of a
scalar field is defined as $H\left(  x^{\mu}\right)  =\phi\left(  x^{\mu
}\right)  -1$ and the action of the scalar field $\phi\left(  x^{\mu}\right)
$ is $S=\frac{m^{2}}{\lambda}%
%TCIMACRO{\dint }%
%BeginExpansion
{\displaystyle\int}
%EndExpansion
dx^{2}\left[  \frac{1}{2}\left(  \partial\phi\right)  ^{2}-\frac{1}{2}\left(
\phi^{2}\left(  x_{0},x\right)  -1\right)  ^{2}\right]  $. In this model, we
consider the second-order fluctuation operator $D_{x}=-\frac{d^{2}}{dx^{2}}+4$
in a finite interval $I=\left[  0,l\right]  $, $l=mL/\sqrt{2}$ with the
Dirichlet boundary condition.

\subsection{The one-loop effective action}

The shifted local regularized one-loop effective action reads%
\begin{equation}
W\left(  s;q;x,y\right)  =-\tilde{\mu}^{2s}\Gamma\left(  s\right)  \frac
{1}{4l}\sum_{n=-\infty}^{\infty}\frac{e^{i\pi n\left(  x-y\right)  /l}-e^{i\pi
n\left(  x+y\right)  /l}}{\left(  n^{2}\pi^{2}/l^{2}+4+q\right)  ^{s}%
}.\label{HiggsWqxy}%
\end{equation}
The sum in eq. (\ref{HiggsWqxy}) can be \textit{exactly} converted into an
integral:%
\begin{align}
W\left(  s;q;x,y\right)   &  =-\frac{\tilde{\mu}^{2s}}{2^{3/2+s}\sqrt{\pi}%
}\int_{0}^{\infty}dt\left(  \frac{\pi}{l}\frac{\sqrt{4+q}}{t}\right)
^{1/2-s}J_{-1/2+s}\left(  \frac{l}{\pi}\sqrt{4+q}t\right) \nonumber\\
&  \times\frac{\left(  \cos\frac{x-y}{t}-\cos\frac{x+y}{t}\right)  \sinh
t}{\left(  \cosh t-\cos\frac{x-y}{t}\right)  \left(  \cosh t-\cos\frac{x+y}%
{t}\right)  }.\label{HiggsWqxyEx}%
\end{align}

The global regularized one-loop effective action is the trace of $W\left(
s;0;x,y\right)  $:%
\begin{align}
W_{s}  &  =TrW\left(  s;0;x,y\right)  =\int_{0}^{l}dxW\left(  s;0;x,x\right)
\nonumber\\
&  =-\frac{1}{2}\tilde{\mu}^{2s}\Gamma\left(  s\right)  \left(  \frac{l}{\pi
}\right)  ^{2s}Z\left(  s;\frac{2l}{\pi}\right)  ,
\end{align}
where $Z\left(  s;a\right)  $ is the Epstein-Hurwitz zeta function \cite{EBK}.

The regularized one-loop effective action without the regularization parameter
$s$ can be achieved by Laurent expanding $W_{s}$ around $s=0$,%
\begin{align}
W_{s} &  =\frac{1}{4s}-\frac{1}{4}\gamma_{E}+l-\frac{1}{2}\ln\frac{2}%
{\tilde{\mu}\left(  1-e^{-4l}\right)  }\nonumber\\
&  +\frac{1}{4}\sum_{p=2}^{\infty}\left\{  \sum_{\beta=0}^{p}\sum_{\alpha
=0}^{\beta}\frac{\left(  -1\right)  ^{\frac{\alpha+\beta}{2}-1}\left(
2^{\beta-\alpha}-2\right)  B_{\beta-\alpha}\pi^{\beta-\alpha}}{\alpha!\left(
\beta-\alpha\right)  !\left(  p-\beta\right)  !}\left[  \left.  \frac
{\partial^{\alpha}\frac{1}{\Gamma\left(  \xi\right)  }}{\partial\xi^{\alpha}%
}\right\vert _{\xi=1}\right]  \left(  \ln\frac{\tilde{\mu}^{2}}{4}\right)
^{p-\beta}\right\}  s^{p-1}\nonumber\\
&  -\frac{l}{2\sqrt{\pi}}\sum_{p=1}^{\infty}\left[  \sum_{\alpha=0}^{p}%
\frac{1}{\alpha!\left(  p-\alpha\right)  !}\left(  \ln\frac{\tilde{\mu}^{2}%
}{4}\right)  ^{p-\alpha}\Gamma^{\left(  \alpha\right)  }\left(  -\frac{1}%
{2}\right)  \right]  s^{p}\nonumber\\
&  -\sqrt{\frac{2l}{\pi}}\sum_{n=1}^{\infty}\frac{1}{\sqrt{n}}\sum
_{p=1}^{\infty}\left[  \sum_{\alpha=0}^{p}\frac{1}{\alpha!\left(
p-\alpha\right)  !}\left(  \ln\frac{nl\tilde{\mu}^{2}}{2}\right)  ^{p-\alpha
}K_{-1/2}^{\left(  \alpha\right)  }\left(  4nl\right)  \right]  s^{p}.
\end{align}
Then taking $s=0$ and dropping the divergent negative power term gives%
\begin{equation}
W=-\frac{1}{4}\gamma_{E}+l+\frac{1}{2}\ln\frac{\left(  1-e^{-4l}\right)
\tilde{\mu}}{2}.
\end{equation}

Moreover, besides the exact expression (\ref{HiggsWqxyEx}), we can also find
an approximate solution but somewhat simple expression for $W\left(
s;q;x,y\right)  $ and $W_{s}$. Replacing approximately the sum $\sum
_{n=-\infty}^{\infty}$ by the integral $\int_{-\infty}^{\infty}dn$ gives
\begin{align}
W\left(  s;q;x,y\right)   &  \simeq-\frac{\tilde{\mu}^{2s}}{2^{s}\sqrt{2\pi}%
}\left[  \left(  \frac{\sqrt{4+q}}{\left\vert x-y\right\vert }\right)
^{1/2-s}K_{1/2-s}\left(  \sqrt{4+q}\left\vert x-y\right\vert \right)  \right.
\nonumber\\
&  -\left.  \left(  \frac{\sqrt{4+q}}{\left\vert x+y\right\vert }\right)
^{1/2-s}K_{1/2-s}\left(  \sqrt{4+q}\left\vert x+y\right\vert \right)  \right]
\end{align}
and%
\begin{equation}
W_{s}\simeq\frac{\tilde{\mu}^{2s}}{2}\frac{\Gamma\left(  s\right)  l^{2s}%
}{\left(  1-2s\right)  \pi^{2s}}\left.  _{2}F_{1}\right.  \left(  -\frac{1}%
{2}+s,s;\frac{1}{2}+s;-\frac{4l^{2}}{\pi^{2}}\right)  .
\end{equation}
Laurent expanding $W_{s}$ around $s=0$, taking $s=0$ and dropping the
divergent part gives%
\begin{equation}
W\simeq1-\frac{\gamma_{E}}{2}+\frac{2l}{\pi}\tan^{-1}\left(  \frac{2l}{\pi
}\right)  -\ln\frac{\sqrt{\pi^{2}/l^{2}+4}}{\tilde{\mu}}.
\end{equation}

\subsection{The vacuum energy}

By a similar treatment, through solving eq. (\ref{EqofveDeq}) with
$D_{x}=-\frac{d^{2}}{dx^{2}}+4$, we can obtain the shifted local regularized
vacuum energy:%
\begin{equation}
E_{0}\left(  \epsilon;q;x,y\right)  =\frac{\tilde{\mu}^{2\epsilon}}{4l}%
\sum_{n=-\infty}^{\infty}\left[  e^{i\pi n\left(  x-y\right)  /l}-e^{i\pi
n\left(  x+y\right)  /l}\right]  \left(  \frac{n^{2}\pi^{2}}{l^{2}%
}+4+q\right)  ^{1/2-\epsilon}.\label{LocalveHiggs}%
\end{equation}
In addition, the sum in eq. (\ref{LocalveHiggs}) can be \textit{exactly}
converted into an integral:%
\begin{align}
E_{0}\left(  \epsilon;q;x,y\right)   &  =\frac{\tilde{\mu}^{2\epsilon}}%
{\Gamma\left(  -1/2+\epsilon\right)  }\frac{1}{2^{1+\epsilon}\sqrt{\pi}%
}\nonumber\\
&  \times\int_{0}^{\infty}dt\left(  \frac{\pi}{l}\frac{\sqrt{4+q}}{t}\right)
^{1-\epsilon}J_{-1+\epsilon}\left(  \frac{l}{\pi}\sqrt{4+q}t\right)
\frac{\left(  \cos\frac{x-y}{t}-\cos\frac{x+y}{t}\right)  \sinh t}{\left(
\cosh t-\cos\frac{x-y}{t}\right)  \left(  \cosh t-\cos\frac{x+y}{t}\right)
}.\label{velocalex}%
\end{align}
The global regularized vacuum energy is the trace of $E_{0}\left(
\epsilon;0;x,y\right)  $:%
\begin{equation}
E_{0}\left(  \epsilon\right)  =\frac{\tilde{\mu}^{2\epsilon}}{2}\left(
\frac{\pi}{l}\right)  ^{1-2\epsilon}Z\left(  \epsilon-\frac{1}{2},\frac
{2l}{\pi}\right)  .
\end{equation}
To remove the divergence, we Laurent expand $E_{0}\left(  \epsilon\right)  $,%
\begin{align}
&  E_{0}\left(  \epsilon\right)  =\frac{l}{2\pi\varepsilon}-\frac{l}{2\pi
}\left(  \frac{\pi}{l}+\ln\frac{1}{\tilde{\mu}^{2}}+1\right)  -\sum
_{n=1}^{\infty}\frac{1}{n\pi}K_{1}\left(  4nl\right) \nonumber\\
&  +\epsilon\left\{  \frac{1}{2}\ln\frac{4}{\tilde{\mu}^{2}}+\frac{l}{2\pi
}\left[  \frac{1}{2}\left(  \ln\frac{1}{\tilde{\mu}^{2}}\right)  ^{2}+\ln
\frac{1}{\tilde{\mu}^{2}}-\frac{\pi^{2}}{6}-1\right]  \right. \nonumber\\
&  \left.  +\sum_{n=1}^{\infty}\left[  \frac{1}{4n^{2}\pi l}K_{0}\left(
4nl\right)  +\frac{1}{n\pi}K_{1}\left(  4nl\right)  \left(  \ln\frac
{1}{2nl\tilde{\mu}^{2}}-\gamma_{E}+2\right)  \right]  \right\}  +\cdots.
\end{align}
Taking $\epsilon=0$ and dropping the divergent negative power term, we arrive
at the regularized vacuum energy without the regularization parameter
$\epsilon$:%
\begin{equation}
E_{0}=-\frac{l}{2\pi}\left(  1+\frac{\pi}{l}+\ln\frac{1}{\tilde{\mu}^{2}%
}\right)  -\sum_{n=1}^{\infty}\frac{1}{n\pi}K_{1}\left(  4nl\right)  .
\end{equation}

Besides, we can also approximately work out the summation in eq.
(\ref{LocalveHiggs}) by replacing the sum $\sum_{n=-\infty}^{\infty}$ with an
integral $\int_{-\infty}^{\infty}dn$. This gives a somewhat simple expression
for $E_{0}\left(  q;x,y\right)  $%
\begin{equation}
E_{0}\left(  \epsilon;q;x,y\right)  \simeq\frac{\tilde{\mu}^{2\epsilon}\left(
4+q\right)  ^{1/2-\epsilon/2}}{2^{\epsilon}\sqrt{\pi}\Gamma\left(
\epsilon-1/2\right)  }\left[  \frac{K_{1-\epsilon}\left(  \sqrt{4+q}\left\vert
x-y\right\vert \right)  }{\left\vert x-y\right\vert ^{1-\epsilon}}%
-\frac{K_{1-\epsilon}\left(  \sqrt{4+q}\left\vert x+y\right\vert \right)
}{\left\vert x+y\right\vert ^{1-\epsilon}}\right]  ,
\end{equation}
which is an asymptotic expression for eq. (\ref{velocalex}) for large $l$. The
global regularized vacuum energy then reads%
\begin{equation}
E_{0}\left(  \epsilon\right)  \simeq\tilde{\mu}^{2\epsilon}\left[  \frac
{l}{4^{\epsilon}\sqrt{\pi}}\frac{\Gamma\left(  -1+\epsilon\right)  }%
{\Gamma\left(  -1/2+\epsilon\right)  }-\frac{1}{4^{1/2+\epsilon}}\right]  .
\end{equation}
An approximate expression of the regularized vacuum energy without the
regularization parameter $\epsilon$ reads%

\begin{equation}
E_{0}\simeq-\frac{l}{2\pi}\left[  \left(  1+\frac{\pi}{l}+\ln\frac{1}%
{\tilde{\mu}^{2}}\right)  \right]  -\frac{1}{4\pi}G_{1,3}^{3,0}\left(
4l^{2}\left\vert
\begin{array}
[c]{c}%
1\\
\frac{1}{2},0,-\frac{1}{2}%
\end{array}
\right.  \right)  .
\end{equation}

\subsection{The counting function and the spectrum}

The solution of eq. (\ref{dieqofN}) gives the local counting function,%

\begin{equation}
N\left(  \lambda;x,y\right)  =\frac{1}{l}\left[  e^{i\pi n\left(  x-y\right)
/l}-e^{i\pi n\left(  x+y\right)  /l}\right]  \sum_{n=1}^{\infty}\theta\left(
\lambda-\left(  \frac{n^{2}\pi^{2}}{l^{2}}+4\right)  \right)  .
\end{equation}
Taking trace gives the global counting function,%
\begin{align}
N\left(  \lambda\right)   &  =TrN\left(  \lambda;x,y\right)  =\int_{0}%
^{l}dx\frac{1}{l}\left(  1-e^{i2\pi nx/l}\right)  \sum_{n=1}^{\infty}%
\theta\left(  \lambda-\left(  \frac{n^{2}\pi^{2}}{l^{2}}+4\right)  \right)
\nonumber\\
&  =\sum_{n=1}^{\infty}\theta\left(  \lambda-\left(  \frac{n^{2}\pi^{2}}%
{l^{2}}+4\right)  \right)  .
\end{align}
From the counting function, one can achieve the eigenvalue spectrum:%
\begin{equation}
\lambda_{n}=\frac{n^{2}\pi^{2}}{l^{2}}+4,\text{ \ }n=1,2,\cdots.
\end{equation}

\section{Conclusions\label{conclusions}}

In this paper, we suggest an approach for calculating one-loop effective
actions, vacuum energies, and spectral counting functions: constructing the
equations for them so that they can be obtained by solving equations.

We solve some exact solutions for one-loop effective actions, vacuum energies,
and spectral counting functions, including a free massive scalar field in
$\mathbb{R}^{n}$, scalar fields in three-dimensional hyperbolic space $H_{3}$
and $H_{3}/Z$, a scalar field in $S^{1}$, and the Higgs model in a
$(1+1)$-dimensional finite interval.

We construct the series expansion for local one-loop effective actions, vacuum
energies, and spectral counting functions. The result can be used to find
approximate solutions. In order to remove the divergence, renormalization
procedures are used.

In our treatment, the physical quantities such as one-loop effective actions,
vacuum energies, and spectral counting functions play the roles as spectral
functions in spectral problems. This suggests us that the physical quantities
like the one-loop effective actions and vacuum energies also can be used as
tools in spectral problems.

\acknowledgments{We are very indebted to Dr. G. Zeitrauman for his
encouragement. This work is supported in part by NSF of China
under Grant No. 10605013.}

\end{document}